\documentclass{nature_modified}
\usepackage[colorlinks=true,urlcolor=blue,linkcolor=blue,citecolor=blue]{hyperref}
\usepackage{graphicx,caption,subcaption}
\usepackage{multibib}
\usepackage{lscape}
\usepackage{xcolor}
\usepackage{float}
\usepackage{amsmath}
\usepackage{amssymb}
\usepackage{longtable}
\usepackage{graphicx}
\makeatletter

\usepackage{orcidlink}
\usepackage{lineno}
\makeatother
\newcommand{\aj}{Astron. J.}

\newcommand{\araa}{Annual Review of Astronomy and Astrophysics}
\newcommand{\apj}{Astrophys. J.}
\newcommand{\apjl}{Astrophys. J., Letters}
\newcommand{\apjs}{Astrophys. J., Suppl. Ser.}
\newcommand{\ao}{Applied Optics}
\newcommand{\apss}{Astrophysics and Space Science}
\newcommand{\aap}{Astron. Astrophys.}

\newcommand{\aaps}{Astron. Astrophys., Suppl. Ser.}

\newcommand{\mnras}{Mon. Not. R. Astron. Soc.}

\newcommand{\pasp}{Publ. Astron. Soc. Pacific}

\newcommand{\ssr}{Space Science Reviews}

\newcommand{\nat}{Nature}

\title{An extended low-density atmosphere around the Jupiter-sized planet WASP-193\,b}
\author{
Khalid Barkaoui \orcidlink{0000-0003-1464-9276}$^{1,2,3,4}$  \thanks{corresponding author ({\color{blue}khalid.barkaoui@uliege.be})},
Francisco J. Pozuelos \orcidlink{0000-0003-1572-7707}$^{5,1}$  \thanks{corresponding author ({\color{blue}pozuelos@iaa.es})},
Coel Hellier$^{6}$,
Barry Smalley \orcidlink{0000-0002-3456-087X}$^{6}$,
Louise D. Nielsen \orcidlink{0000-0002-5254-2499}$^{7,8}$,
Prajwal Niraula \orcidlink{0000-0002-8052-3893}$^{3}$,
Micha\"{e}l Gillon \orcidlink{0000-0003-1462-7739}$^{1}$, 
Julien de Wit \orcidlink{0000-0003-2415-2191}$^{3}$,
Simon M\"{u}ller \orcidlink{0000-0002-8278-8377}$^{9}$,
Caroline Dorn \orcidlink{0000-0001-6110-4610}$^{9}$, 
Ravit Helled$^{9}$, 
Emmanuel Jehin \orcidlink{0000-0001-8923-488X}$^{10}$, 
Brice-Olivier Demory \orcidlink{0000-0002-9355-5165}$^{11}$, 
V. Van Grootel \orcidlink{0000-0003-2144-4316}$^{10}$,
Abderahmane Soubkiou$^{2,12}$,
Mourad Ghachoui$^{1,2}$,
David. R. Anderson \orcidlink{0000-0001-7416-7522}$^{13}$, 
Zouhair Benkhaldoun \orcidlink{0000-0001-6285-9847}$^{2}$,
Francois Bouchy \orcidlink{0000-0002-7613-393X}$^{7}$,
Artem Burdanov \orcidlink{0000-0001-9892-2406}$^{3}$, 
Laetitia Delrez \orcidlink{0000-0001-6108-4808}$^{1,10}$,
Elsa Ducrot \orcidlink{0000-0002-7008-6888}$^{14,15}$, 
Lionel Garcia \orcidlink{0000-0002-4296-2246}$^{1,16}$, 
Abdelhadi Jabiri$^{2}$, 
Monika Lendl \orcidlink{0000-0001-9699-1459}$^{7}$, 
Pierre F. L. Maxted \orcidlink{0000-0003-3794-1317}$^{6}$, 
Catriona A. Murray \orcidlink{0000-0001-8504-5862}$^{17}$, 
Peter Pihlmann Pedersen$^{18}$, 
Didier Queloz \orcidlink{0000-0002-3012-0316}$^{7,18}$,
Daniel Sebastian \orcidlink{0000-0002-2214-9258}$^{19}$,
Oliver Turner \orcidlink{0000-0002-8216-2796}$^{7,20}$,
Stephane Udry$^{7}$,
Mathilde Timmermans \orcidlink{0009-0008-2214-5039}$^{1}$, 
Amaury H. M. J. Triaud \orcidlink{0000-0002-5510-8751}$^{19}$,
and Richard G. West$^{13,21}$
}

\begin{document} 

\maketitle

\begin{affiliations}
\begin{footnotesize}
\item Astrobiology Research Unit, Universit\'e de Li\`ege, 19C All\'ee du 6 Ao\^ut, 4000 Li\`ege, Belgium 
\item Oukaimeden Observatory, High Energy Physics and Astrophysics Laboratory, Faculty of sciences Semlalia, Cadi Ayyad University, Marrakech, Morocco 
\item Department of Earth, Atmospheric and Planetary Science, Massachusetts Institute of Technology, 77 Massachusetts Avenue, Cambridge, MA 02139, USA
\item Instituto de Astrof\'isica de Canarias (IAC), Calle V\'ia L\'actea s/n, 38200, La Laguna, Tenerife, Spain 
\item Instituto de Astrof\'isica de Andaluc\'ia (IAA-CSIC), Glorieta de la Astronom\'ia s/n, 18008 Granada, Spain 
\item Astrophysics Group, Keele University, Staffordshire, ST5 5BG, UK 
\item Observatoire de Gen\`eve, Universit\'e de Gen\`eve, Chemin Pegasi 51, 1290 Sauverny, Switzerland 
\item European Southern Observatory, Karl-Schwarzschildstr. 2, D-85748 Garching bei M{\"u}nchen, Germany 
\item University of Zurich, Institute of Computational Sciences, Winterthurerstrasse 190, CH-8057 Zurich, Switzerland  
\item Space sciences, Technologies and Astrophysics Research (STAR) Institute, Universit\'e de Li\`ege, Belgium
\item Center for Space and Habitability, University of Bern, Gesellschaftsstrasse 6, CH-3012, Bern, Switzerland 
\item Instituto de Astrofisica e Ciencias do Espaco, Universidade do porto, CAUP, Rua das Estrelas, 150-762 Porto, Portugal 
\item Department of Physics, University of Warwick, Gibbet Hill Road, Coventry CV4 7AL, UK 
\item Paris Region Fellow, Marie Sklodowska-Curie Action
\item AIM, CEA, CNRS, Universit\'e Paris-Saclay, Universit\'e de Paris, F-91191 Gif-sur-Yvette, France 
\item Center for Computational Astrophysics, Flatiron Institute, New York, NY, USA
\item Department of Astrophysical \& Planetary Sciences, University of Colorado Boulder, 2000 Colorado Avenue, Boulder, CO 80309, USA 
\item Cavendish Laboratory, JJ Thomson Avenue, Cambridge, CB3 0H3, UK 
\item School of Physics \& Astronomy, University of Birmingham, Edgbaston, Birmingham, B15 2TT, UK 
\item Space Forge, Unit 10, Eastgate Business Park, Wentloog Ave, Rumney, Cardiff, CF3 2EY 
\item Centre for Exoplanets and Habitability, University of Warwick, Gibbet Hill Road, Coventry CV4 7AL, UK 
\end{footnotesize}
\end{affiliations}

%- Bold paragraph (~200 words)
\begin{abstract}
Gas giants transiting bright nearby stars provide crucial insights into planetary system formation and evolution mechanisms.  Most of these planets exhibit certain average characteristics, serving as benchmarks for our understanding of planetary systems. However, outliers like the planet we present in this study, WASP-193\,b, offer unique opportunities to explore unconventional formation and evolution processes. This planet completes an orbit around its $V_{\rm mag} = 12.2$ F9 main-sequence host star every 6.25~d. Our analyses found that WASP-193\,b has a mass of $M_p = 0.139 \pm 0.029 ~M_{\rm Jup}$ and a radius of $R_p = 1.464 \pm 0.058~R_{\rm Jup}$, translating into an extremely low density of $\rho_p = 0.059 \pm 0.014~{\rm g~cm^{-3}}$, at least one order of magnitude less than standard gas giants like Jupiter. Typical gas giants such as Jupiter have densities that range between 0.2 and 2 ${\rm g~cm^{-3}}$. The combination of its large transit depth ($dF\sim 1.4$~\%), its extremely-low density, its high-equilibrium temperature ($T_{\rm eq} = 1254 \pm 31~K$), and the infrared brightness of its host star (magnitude $K_{\rm mag}=10.7$) makes WASP-193\,b an exquisite target for characterization by transmission spectroscopy (transmission spectroscopy metric: $TSM\sim600$). One single JWST transit observation would yield detailed insights into its atmospheric properties and planetary mass, providing a unique window to explore the mechanisms behind its exceptionally low density and shed light on giant planets' diverse nature.

\end{abstract}

\noindent
WASP-193 was observed by the WASP-South\cite{Pollacco2006} survey in 2006--2008 and 2011--2012. The WASP photometry revealed a periodic signal consistent with transits of a giant planet with an orbital period of 6.25d (see Methods).

 \noindent
 We used TRAPPIST-South\cite{Gillon2011} and SPECULOOS-South\cite{Jehin2018Msngr} ground-based telescopes to observe five transits of WASP-193\,b, one in the $I+z$ filter (02 January 2015), one in a `Blue-Blocking' ($BB$) filter (06 April 2015), one in the Sloan-$z'$ filter (27 January 2017), and two in the Johnson-$B$ filter (08 June 2019). Additional photometric measurements were collected by the \emph{TESS}\cite{Ricker2015} mission with 1800, 600, and 120~s cadences, of which we used in our analysis the PCDSAP fluxes corresponding to 600~s (Sector 36 in 2021) and 120~s (Sector 63 in 2023) (see Methods). In addition, we used the Euler-1.2m/CORALIE\cite{Queloz2000} and ESO-3.6m/HARPS\cite{Mayor2003Msngr} spectrographs to observe WASP-193. We obtained 12 and 11 spectroscopic measurements with CORALIE and HARPS in 2015 and 2019, respectively (see Methods).

\begin{figure*}
	\centering
	\includegraphics[scale=0.35]{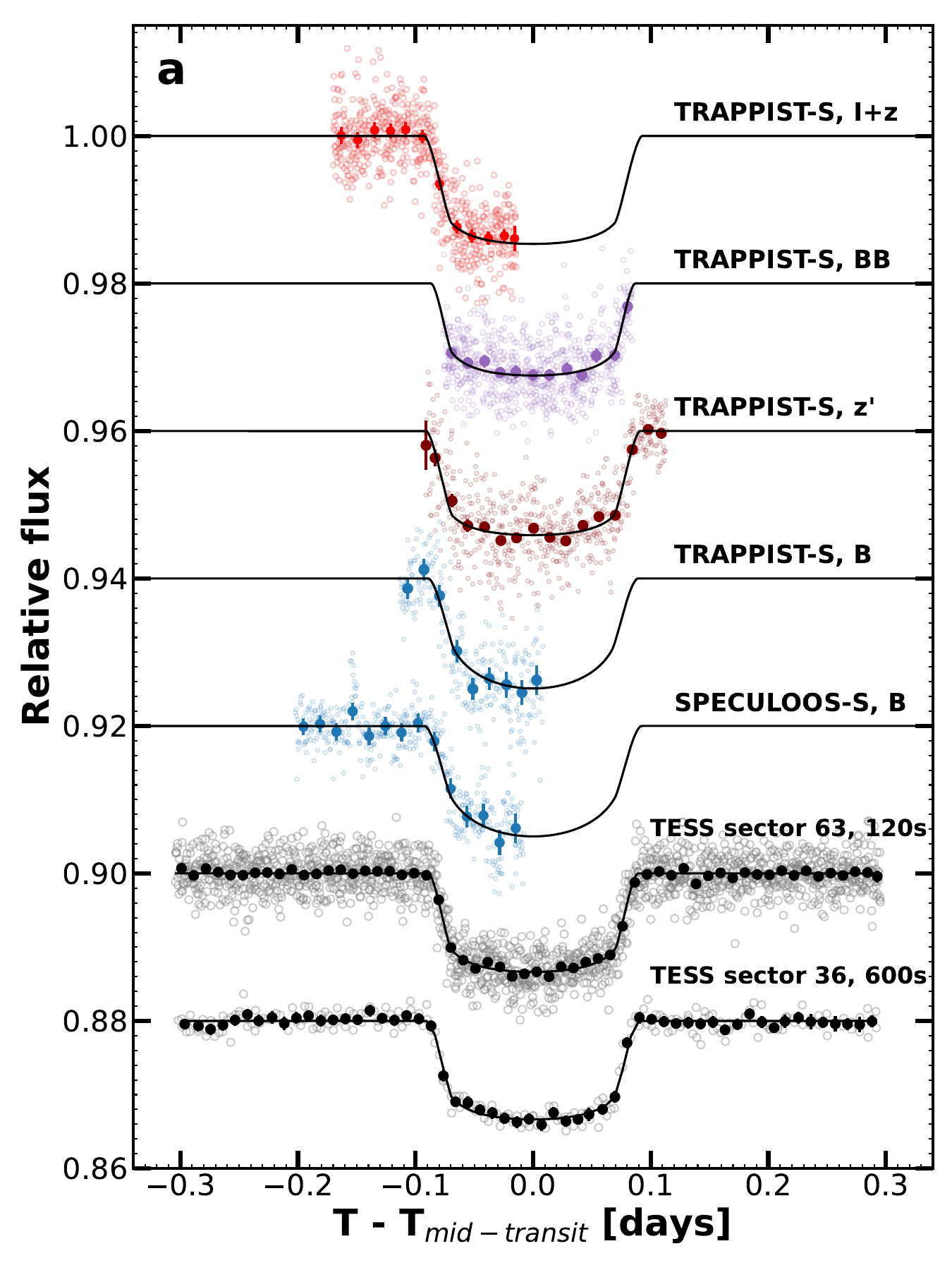}
	\includegraphics[scale=0.35]{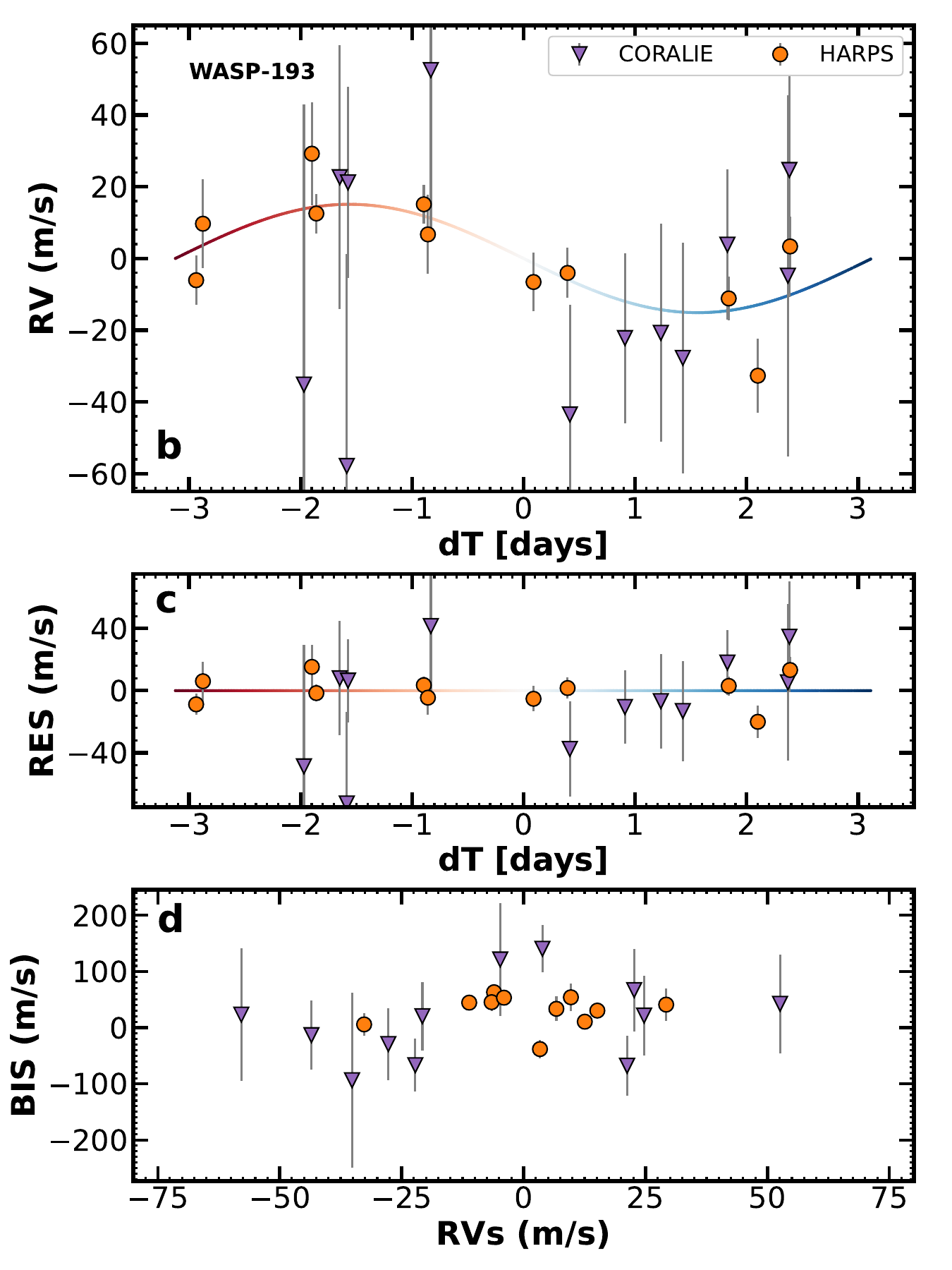}
	\caption{ Photometric and spectroscopic follow-up.  {\bf a}, Follow-up transit photometry of WASP-193\,b. Four transits were observed by TRAPPIST-South, one by SPECULOOS-South, and the observations performed by \emph{TESS} in Sector 36 and 63 with 10 and 2-min cadences, respectively. In all the cases, the period-folded is displayed using the best-fitting transit ephemeris deduced from our global MCMC analysis. Each transit light curve has been corrected by the baseline model presented in Supplementary Table~1. The coloured points with error bars are data binned to 20~min, and the best-fitting model is superimposed in black.  The light curves are shifted by 0.02 along the $y$-axis for clarity.
    {\bf b},  RVs obtained with the CORALIE and HARPS spectrographs for WASP-193 period-folded using the best-fitting orbital model, which is superimposed as a  solid line (in $m~s^{-1}$). RVs color line of the best fitting shows the Doppler effect. 
    {\bf c},  RV residuals of the fit (in $m~s^{-1}$). The coloured triangles and circles correspond to the CORALIE and the HARPS data, respectively. 
    {\bf d}, Bisector-span as a function of RV (in $m~s^{-1}$)  of WASP-193 obtained with the CORALIE and HARPS spectrographs. We assumed that the bisector-span error bars as twice the RV error bars\cite{Maxted_2016AA}.
    %\cite{Maxted_2016AA,Basturk_2011AA,Martinez_2005AA}.
 }
	\label{Photometry_RVs}
\end{figure*}

\noindent
All the follow-up data were consistent with the planetary origin of the transit signal identified by WASP. The high-precision follow-up light curves acquired in different filters showed no chromatic dependence (see Fig.~\ref{Photometry_RVs}, panel $a$). Indeed, the transit depths obtained with different bands agree at 1$\sigma$ level (see Methods). The radial velocities phased up with the transit time and period of 6.25\,d, consistent with the transit timings, and the bisector analysis of the HARPS and CORALIE spectra\cite{Torres_2004} did not reveal any dependence of the measured radial velocities with the bisector span 
(see Fig.~\ref{Photometry_RVs}, panels $b$, $c$, and $d$). 

\noindent
To constrain the stellar atmospheric parameters of WASP-193, we co-added HARPS and CORALIE spectra to produce a single stellar spectrum with a high signal-to-noise ratio (S/N$\sim$75) and analyzed it as described in ref.\cite{Doyle2013}. This analysis enabled us to constrain the following parameters: the spectral type, the metallicity $[Fe/H]$, the effective temperature $T_{\rm eff}$, the surface gravity $\log g_\star$, and the rotational velocity $V\sin(i)$ (see Methods). The resulting values show that WASP-193 is a F9-type Sun-like star with $T_{\rm eff} = 6076 \pm 120$~K, $\log g_\star = 4.1 \pm 0.1$~dex, $[Fe/H] = -0.06 \pm 0.09$~dex and $V\sin (i) = 4.3 \pm 0.8$~km~s$^{-1}$. Using our measured $T_{\rm eff}$, the star's magnitudes in different broadband filters, and its Gaia eDR3 parallax\cite{Gaia_Collaboration_2020_DR3}, we conducted a SED-fitting to derive a stellar radius of $R_\star = 1.225_{-0.029}^{+0.032}~R_{\odot}$ by employing the {\it EXOFASTv2\cite{Eastman_2019PASP}} pipeline (see Methods). Adopting $T_{\rm eff}$, [Fe/H], and $R_\star$ as the basic input set, we then derived the stellar mass $M_\star$ and age $t_\star$ from stellar evolutionary modeling. We used the {\it CLES} (Code Liégeois d’Évolution Stellaire\cite{Scuflaire_2008_CLES}), building the best-fit stellar evolutionary track according to the input parameters following the Levenberg-Marquadt minimisation scheme, as explained in ref.\cite{Salmon_2021_Stellar_evol}. We obtained $M_\star=1.102\pm0.070 ~M_{\odot}$ and $t_\star=4.43\pm1.92$~Gyr (see Methods).
% \autoref{stellarpar} summarizes the stellar parameters inferred from our analyses.
%different spectroscopic analyses.

\noindent
%%%%%%%%%%%%% MCMC analysis
We derived the system's parameters by conducting a global model fitting together the photometric and spectroscopic data using the latest version of the Markov Chain Monte Carlo (MCMC) code described in ref.\cite{Gillon2012}, employing the Mandel \& Agol \cite{Mandel2002} and a 2-body Keplerian models\cite{Murray2010}, respectively (see Methods). 
Table~\ref{mcmcwasp193b} shows the deduced stellar and planetary physical parameters along with their 1$\sigma$ intervals. 

\begin{figure*}[!ht]
	\centering
	\includegraphics[scale=0.45]{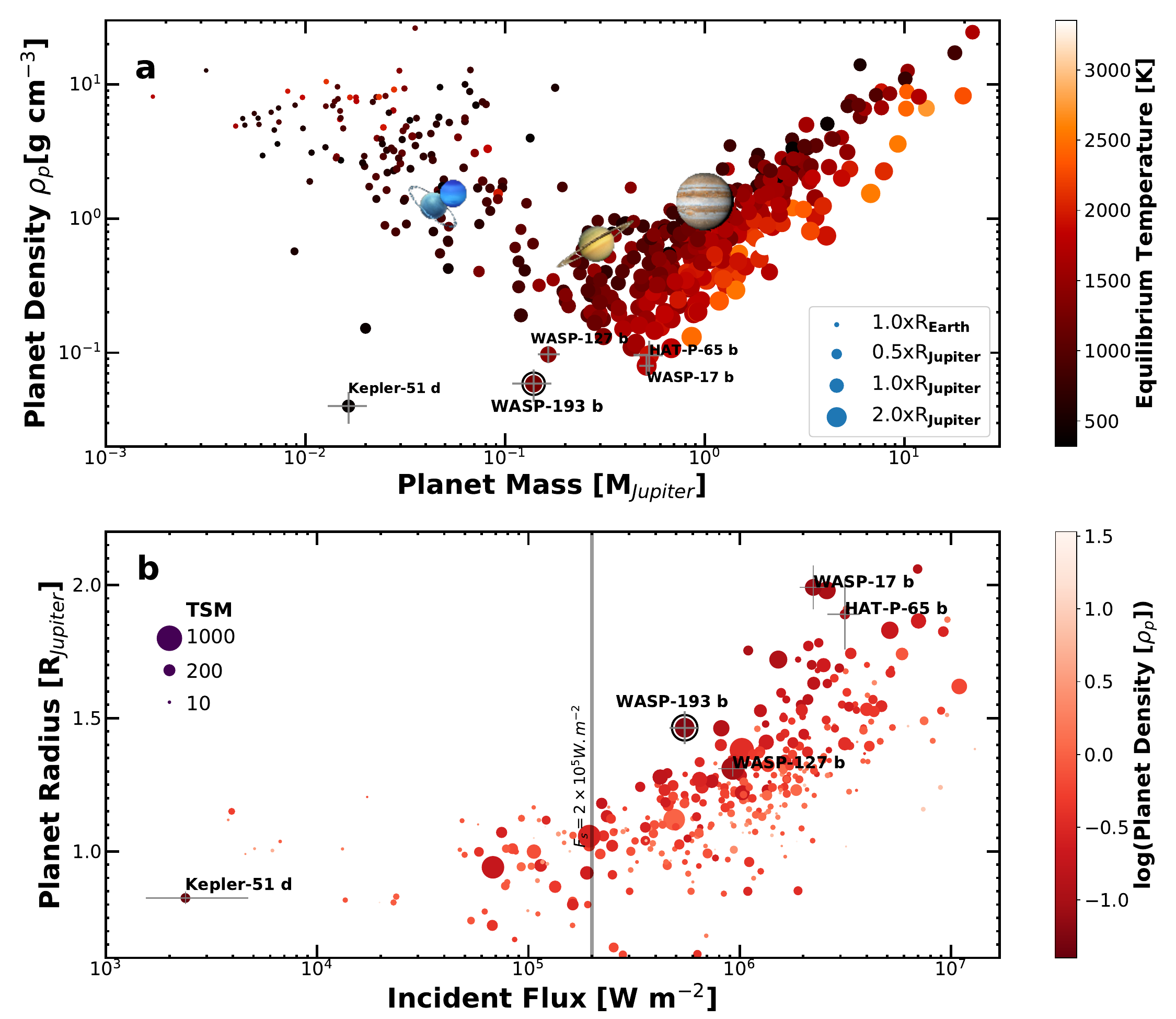}
	\caption{ 
Planetary diagrams of known transiting exoplanets with radius and mass precisions better than 8\% and 25\%, respectively. The data have been extracted from the \texttt{NASA Exoplanets Archive}. 
{\bf a}, Planetary density as a function of the planetary mass. The size of the points scale with the planetary radius. The points are coloured according to their equilibrium temperature. WASP-193\,b and the least dense planets known to date are labelled and displayed with their 1$\sigma$ uncertainties (Kepler-51\,d\cite{Jontof-Hutter_Kepler-51_2022}, WASP-17\,b\cite{Anderson_2011MNRAS_WASP-17b}, WASP-127\,b\cite{Lam_2017AA_WASP-127b} and HAP-P-65\,b\cite{Hartman_2016AJ_HAT-P-65b}).
{\bf b}, Planetary radius as a function of the stellar incident flux. The size of the points scale with the TSM (transmission spectroscopy metric; \cite{Kempton_2018PASP}). The points are coloured according to their planetary density. The radii begin to show a correlation with stellar incident fluxes at $\sim 2\times10^5$~W~m$^{-2}$ (vertical gray line)\cite{Demory_2011}.
}
	\label{wasp193b_Mp_Rhop_Rp_Irrad}
\end{figure*}

\noindent
 We find that planet has a radius of $R_p = 1.464 \pm 0.058~R_{\rm Jup}$ and a mass of $M_p = 0.139\pm 0.029~M_{\rm Jup}$ ($\sim3~M_{\rm Neptune}$), resulting in a surprisingly low density of $\rho_p = 0.059\pm 0.014 ~{\rm g/cm^{3}}$ (see top panel in 
 Fig.~\ref{wasp193b_Mp_Rhop_Rp_Irrad}). This extremely low density positioned WASP-193\,b as the second lightest planet discovered so far after Kepler~51\,d ($\rho = 0.04\pm 0.01~{\rm g/cm^{3}}$) (both planets' densities connected at 1$\sigma$ level), and as the lightest planet over the hot-Jupiter population. It is worth noting that we restricted our comparison to planets that are precisely characterized, with dynamical masses and radii precisions better than 25\% and 8\%, repectively\cite{Luque_2022Sci}. This approach aimed to eliminate any potential bias introduced by comparing WASP-193\,b with inadequately characterized planets that could impede conclusive findings. Notably, these poorly characterized planets exhibit a 1$\sigma$ level association with distinct planet populations, further emphasizing the challenge of performing a meaningful analysis of their nature due to significant degeneracies.

\noindent  
Remarkably, considering the irradiation of about $S_p = 6\times10^5~W/m^2$ (or 440$\times$ larger than that of Earth) inferred from our analysis (see the panel $b$ in Fig.~\ref{wasp193b_Mp_Rhop_Rp_Irrad}), classical evolution models of irradiated giant planets\cite{Fortney2007,Baraffe_2008A,Weiss2013,Sestovic_2018AA} are not able to reproduce WASP-193\,b's radius. Indeed, using the irradiated gas giant model from ref.\cite{Fortney2007} for a core mass of 0 to 10~$M_\oplus$ and age between 1.0 to 4.5 Gyrs,  the predicted radius is found to be 0.9--1.1~$R_{\rm Jup}$. We also used the equation derived by ref.\cite{Weiss2013} from 35 exoplanets with $M_p < 150 ~M_\oplus$. In this case, the resulting predicted radius is found to be $R_p = 0.82 \pm 0.14~R_{\rm Jup}$. Finally, we used the model of ref.\cite{Sestovic_2018AA} to calculate the radius of the planet using data from 286 hot Jupiters with known masses and radii. In this scenario, the predicted radius measures $R_p = 1.1 \pm 0.1~R_{\rm Jup}$. Therefore, WASP-193\,b joins thus the sub-group of `anomalously large' irradiated gas giants\cite{Lopez_2016}.

%%%%% from CD, RH and SM
\noindent
{\bf Planet Interior}\\
\noindent
In the core accretion model, planets at this mass range are expected to have non-negligible amounts of heavy elements\cite{2014prpl.conf..643H}. This is at odds with the low density of WASP-193\,b. Alternatively, the planet could have formed by disk instability, as it was recently shown by ref.\cite{deng2021formation} that magnetically controlled disk fragmentation could lead to intermediate-mass clumps of H-He. The formation mechanism of planets like WASP-193\,b remains uncertain, including its potential orbital evolution and we hope that future formation models can address its formation mechanism.

\begin{figure*}[!ht]
\centering
    \includegraphics[width=.6\linewidth,trim=0cm 0cm 0cm 0cm, clip]{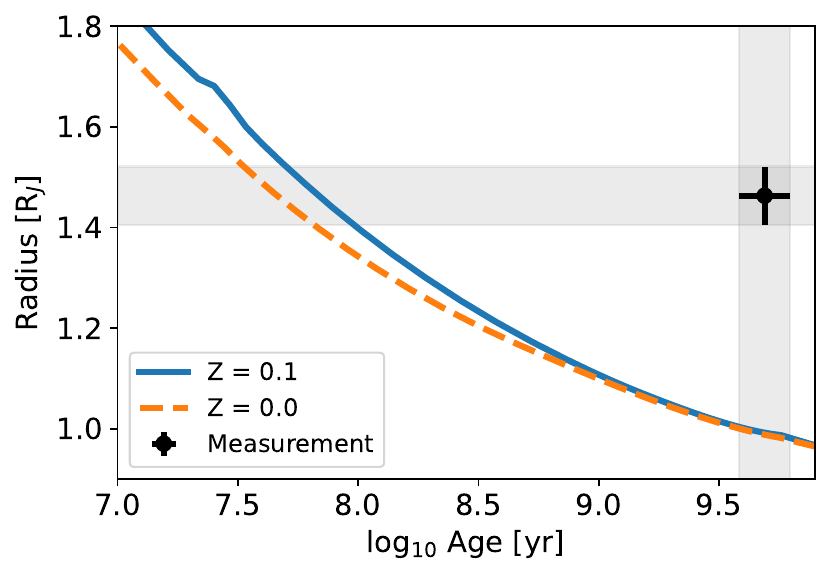}
    \caption{Planetary Radius as a function of age for two evolution models. The orange dashed line and solid blue lines are for $Z = 0.0$ and $0.1$, respectively,  where $Z$ is the planetary metallicity. The radius and age inferred from observations are shown as the black error bars and the grey-shaded regions. The radius evolution of the planet includes the effect of the evolving stellar irradiation (see text for details). Note that the slow cooling of the model with $Z = 0.1$ allows for larger radii compared to a pure H-He composition.}
    \label{fig:wasp193b_evolution}
\end{figure*}

\noindent
We calculated the evolution of WASP-193\,b following the framework presented by ref.\cite{muller2021synthetic}, which assumes a \textit{hot start} core accretion formation scenario\cite{2007ApJ...655..541M}. The models use the equations of state\cite{chabrier2019new} for H-He and for a 50-50 water-rock mix\cite{more1988new}. The planetary radius corresponds to the photosphere where the optical depth drops below 2/3. Gas opacities are taken from ref.\cite{freedman2014gaseous}. Additional details on the interior evolution model can be found in ref.\cite{muller2021synthetic}.
We adapted the evolution models of ref.\cite{muller2021synthetic} to account for the evolution of the host star where the stellar luminosity follows a pre-calculated stellar evolution track. The luminosity as a function of time was calculated with the MESA Isochrones and Stellar Tracks (MIST)\cite{Choi:2016,Dotter:2016} with the mass of $M_\star = 1.068~M_\odot$ and slightly sub-solar metallicity $[Fe/H] = -0.06$ obtained from our spectroscopic analysis. The age-range of the planet was defined by the ages at which the stellar luminosity from the evolution tracks matched the inferred one $L = 1.87^{+0.18}_{-0.16} \, L_\odot$ (see Methods).
\noindent
As there is no clear correlation between stellar metallicity and the composition of giant planets \cite{2019AJ....158..239T}, the bulk composition of WASP-193\,b cannot be inferred from the measured stellar $[Fe/H]$. As a result, we considered two end members of the planetary composition (i.e., planetary metallicity). First, we calculated the evolution of WASP-193\,b  with $Z = 0.0$, representing the lowest metallicity case. Naively, one would expect a pure H-He planet to yield the largest radius (and lowest density) at a given time. However, since a metal-enriched atmosphere has a higher opacity and therefore delays the cooling, a planet with  $Z > 0$ could actually contract more slowly\cite{muller2020challenge}. It has been shown that a turn-over value of $Z$ exists for which the effect of the increased density wins over the delayed cooling. After that, the planetary size decreases with increasing $Z$ (see ref.\cite{muller2020challenge} for further details). For planet WASP-193\,b, the turn-over value is $Z = 0.1$. Therefore, we also considered the evolution of WASP-193\,b with $Z = 0.1$.
The evolution of the planetary radius for the two cases is shown in Fig.~\ref{fig:wasp193b_evolution}. While the enriched atmosphere in the $Z = 0.1$ case yielded a larger radius until around 1 Gyr, it is clear that neither model can reproduce the observed radius of WASP-193\,b for the possible stellar ages. The influence of the stellar evolution is more profound at early times when the star has a higher luminosity. The planet cools down as the stellar luminosity decreases and the star reaches the main sequence.  We find that the planet has a size comparable to that of  WASP-193\,b only for an age of a few 10 Myr, which is inconsistent with the estimated age of the host star.
Clearly, the large observed radius of $\sim$1.5~$R_{\rm Jup}$ at an age of several Gyr cannot be reproduced unless other mechanisms are at work, these include processes that (1) delay the cooling of the planet, or (2) deposit heat deep in the planetary interior and mass loss, all of which are in principle dependent on metallicity $Z$. We discuss such mechanisms below.

\noindent
\paragraph{\bf \normalsize Delay in planetary cooling}
Inefficient cooling can be caused by various processes, such as an enhanced atmospheric opacity that leads to longer cooling time scales, 
interiors that are not fully convective, and phase separations. 
For the former, as we show in Fig.~\ref{fig:wasp193b_evolution}, the effect of prolonged cooling via opacity enhancement is insufficient to explain WASP-193\,b's large radius. Additionally, it would be possible that the opacity is enhanced due to dust grains or cloud decks, which were not included in our models. However, it is unlikely that this would increase the radius beyond a few percent\cite{Poser2019}.
%\cite{2013MNRAS.434.3283V,Poser2019}
Also, in the case of a not fully convective interior, the planet contracts at a lower rate. This can occur when the interior consists of boundary layers and/or composition gradients\cite{vazan2018jupiter}.
%\cite{chabrier2007heat, helled2017fuzziness, vazan2018jupiter}.
Such a configuration can lead to much higher internal temperatures at a given age\cite{leconte2012new} and is expected to lead to an increase in radius by up to $\sim$10\%\cite{kurokawa2015radius}. Therefore, this scenario also cannot explain the large observed radius.
Phase separation could also lead to slower cooling. In giant planets, He-rain can delay the planetary cooling\cite{stevenson1977dynamics}. Its ability to increase the planetary radii is, however, limited to a few percent\cite{fortney2004effects}. More importantly, a planet of $\sim$0.14~$M_{\rm Jup}$ as WASP-193\,b is not expected to reach internal pressures that are high enough for He to become immiscible\cite{fortney2004effects}. 
A self-consistent model that incorporates layered-convection, tidal heating, boundary layers as well as enhanced opacities and dust-rich atmospheres to delay the cooling to its maximum would be required to fully evaluate whether the combined mechanisms are sufficient to explain the measured radius. However, the observed radius of WASP-193\,b is 40-50\% larger than predictions from evolution models. Therefore, these mechanisms are unlikely to add up to account for the inflated radius fully.

\noindent
\paragraph{\bf \normalsize Heat deposition and mass loss}
 The planet may be heated at depth due to Ohmic dissipation or tidal heating. 
However, tidal heating is unlikely to be the mechanism that inflates hot Jupiters since it has been observed that the inflated radii correlate with stellar flux better than with the distance to the hot star\cite{Weiss2013,2011ApJ...729L...7L}.
Currently, Ohmic dissipation seems to be the most promising mechanism\cite{fortney2021hot}. This is because the atmosphere of WASP-193\,b is sufficiently hot ($T_{\rm eq}\sim1250~K$) that trace elements in the H-He gas can be partially ionized and allow for atmospheric currents to penetrate deep into the interior. For hot Jupiters, the radius inflation power depends on the incident flux\cite{Thorngren_2018AJ}. If WASP-193\,b experiences similar inflation mechanisms, this suggests that mechanisms that are coupled to stellar heating, including Ohmic dissipation, are favorable, and possibly, multiple mechanisms are at play. 
Additional heating in the deep interior could be occurring together with  semi-convective and/or boundary  layers. This could effectively trap some of the deposited energy and delay the planetary cooling.\\
We estimate the additional energy required to explain the observed radius of WASP-193\,b independent of the heating mechanism.  We determine the planetary total internal energy $E_0$ at a time $t_0$ where the planet's radius matches the observed one. The energy required for the additional inflation is then $\Delta E = E_0 - E_{\textrm{today}}$, where $E_{\textrm{today}}$ is the internal energy at the mean stellar age. We find that $t_0 = 40$ Myr ($Z = 0$) and $60$ Myr ($Z = 0.1$), which yielded a similar $\Delta E \sim 10^{40}$ erg for the two cases. For comparison, the energy WASP-193\,b receives today by the stellar irradiation is $\sim 5 \times 10^{36}$ erg/yr, about four orders of magnitude lower.

\noindent
Finally, the relatively low mass of WASP-193\,b, its high stellar irradiation, and its location at the edge of the sub-Jovian valley in the M-R diagram might hint that it may be losing mass. If this is the case, the extended observed radius may be related to material evaporating and leaving the planet. However, this is only a speculation, and we stress that there is no indication in the measurements for atmospheric loss. 
It is possible to use theoretical models of mass loss for highly irradiated planets to estimate whether atmospheric escape is significant for WASP-193\,b. We calculate the restricted Jeans escape parameter $\Lambda \propto M_p/T_{eq} R_p$ from ref.\cite{2017A&A...598A..90F}, which can be used to identify planets for which atmospheric escape is important. For WASP-193\,b we find $\Lambda = 3.7$. It was shown in ref.\cite{2017A&A...598A..90F} that $\Lambda$  values lower than the critical value of $\Lambda_T = 15 - 35$ correspond to significant atmospheric loss. As a result, the atmospheric mass loss could be important for WASP-193\,b. However, we emphasize that detailed hydrodynamic simulations should be performed to fully assess the importance of atmospheric escape for WASP-193\,b.\\\\
\noindent
Overall, the evolution model cannot explain the observations of WASP-193\,b, despite its being state-of-the-art.
A self-consistent model that incorporates layered convection, boundary layers, as well as enhanced opacities, and dust-rich atmospheres to delay the cooling to its maximum would be required to evaluate whether the combined mechanisms are sufficient to explain the measured radius. However, this would require considerable effort in improving existing evolution models, which is out of the scope of this study. 
Besides advances in modelling, additional constraints on the atmospheric escape, composition, and temperature structure might be obtained by conducting transit spectroscopy using the JWST, providing key information to unravel the nature of this low-density planet, as discussed in the following.

\begin{figure*}
\centering
        \includegraphics[width=0.9\textwidth]{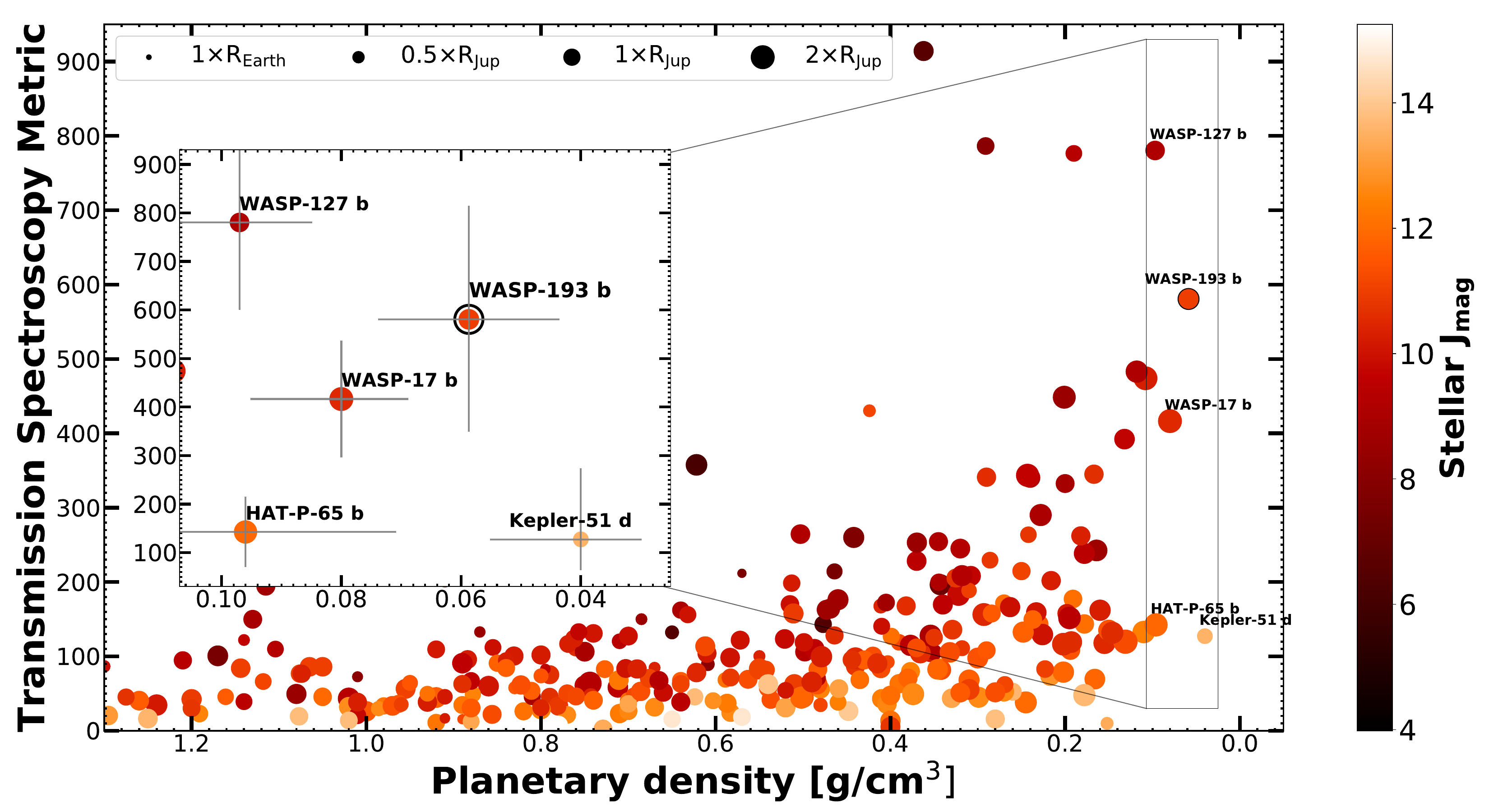}
        \caption{\label{fig:TSM_all}   Feasibility of WASP-193\,b for transmission spectroscopy studies. The transmission spectroscopy metric (TSM) is presented as a function of the planetary density of known transiting exoplanets obtained from the NASA Exoplanet Archive (\url{https://exoplanetarchive.ipac.caltech.edu/}) with radius and mass precisions better than 8\% and 25\%, respectively. The point size scales with the planetary radius. The points are colored according to the host star $J_{\rm mag}$. WASP-193\,b is highlighted with a black contour.
        } 
\end{figure*}

\begin{figure*}
\centering
        \includegraphics[trim={0.3cm 0.25cm 0.5cm 0.25cm},clip, width=0.88\textwidth]{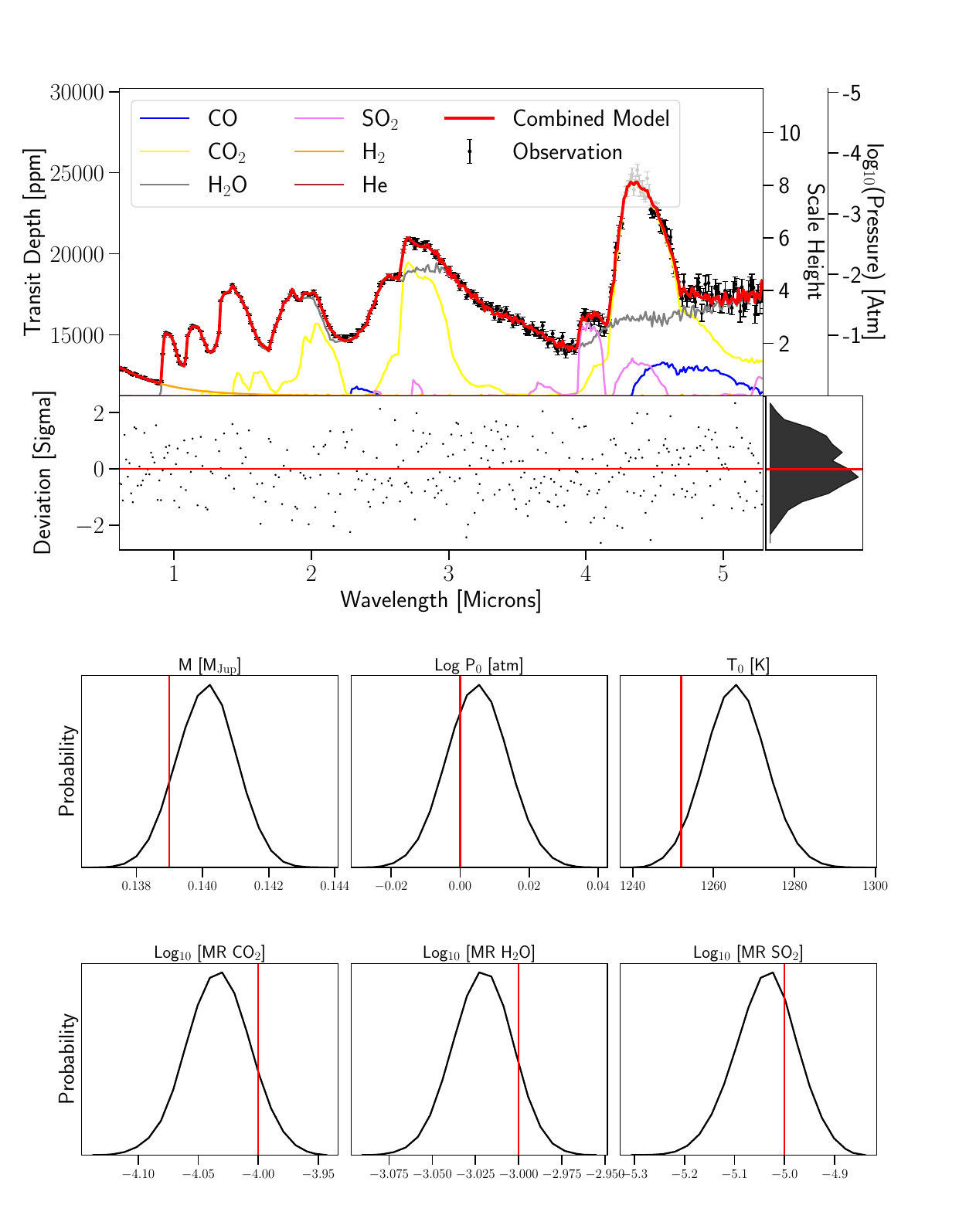}
        \caption{ WASP-193~b as an exquisite target for atmospheric characterization.
        \label{fig:Synth_Spectra} \textbf{a, } Synthetic transmission spectrum model and best-fit model for the injection-retrieval test for a single JWST/NIRSpec/Prism transit of WASP-193\,b. Models and tests are produced with \texttt{TIERRA} pipeline\cite{Niraula_2022NatAs}.  \textbf{b, } Posterior probability distributions of the key planetary properties retrieved showcasing that one single JWST visit can yield the constraints on the main atmospheric properties and the planetary mass within $\sim$0.1 dex and $\sim$1\%, respectively. The red line represents the true value used in generating the synthetic spectrum.
        }
\end{figure*}

\noindent
%%%%%%%%%%%% Atmospheric characterization  prospects
{\bf Atmospheric characterization  prospects} \\
 Owing to its remarkably low density, the hot Jupiter WASP-193\,b is a prime target for atmospheric characterization. As highlighted in Fig.~\ref{fig:TSM_all}, its transmission spectroscopy metric ($TSM$\cite{Kempton_2018PASP}) is amongst the highest to date ($\sim$600). We used the \texttt{TIERRA} retrieval framework\cite{Niraula_2022NatAs} to assess quantitatively the insights that would be accessible with a single visit of \emph{JWST}. To this end, we performed an injection-retrieval on a synthetic WASP-193\,b atmosphere using abundances consistent with WASP-39\,b (see Methods). We assume the use of NIRSpec/Prism as its wavelength coverage provides the best trade-off for atmospheric exploration. The synthetic data, the best fit, and the retrieved planetary parameters are presented in Fig.~\ref{fig:Synth_Spectra} . 
We find that a single transit observation can yield the abundances of strong absorbers within 0.1~dex, assuming no significant opacity challenge\cite{Niraula_2022NatAs} as expected for hot Jupiters\cite{Niraula_2023arXiv230303383N}. Most importantly, we find that the planetary mass is  tightly constrained within $\sim$1\% in comparison to the current $\sim$20\% precision achieved with radial velocity measurements. This tight constraint is enabled by the independent constraints on the atmospheric scale height ($H = 3008 \pm 119$~km), temperature, and abundance accessible via transmission spectroscopy at such high TSM\cite{deWit2013}.
While a single visit with JWST/NIRSpec/Prism would provide the most insights into the planetary atmosphere, a single JWST/NIRISS SOSS observation would also allow to detect the He absorption triplet, this would provide preliminary insights into the current atmospheric loss\cite{seager2000}. 
%\cite{seager2000,spake2018}. 
Ground-based high-resolution measures could later complement this picture by constraining the line profile\cite{allart2018}, helping to validate or dismiss the hypothesis that atmospheric loss is responsible for the observed low-density nature of this planet. Moreover, using the NIRISS/SOSS, it would also be possible to place some constraints on the atmospheric metallicity (M/H), carbon-to-oxygen ratio (C/O), and potassium-to-oxygen ratio (K/O)\cite{feinstein2023}, shedding some light into the formation process of WASP-193\,b. For example, the hypothetical combination of a super-solar metallicity, super-solar K/O ratio, and sub-solar C/O ratio may suggest that the planet formed beyond the H$_{2}$O snow line followed by inward migration, producing an efficient accretion of planetesimals\cite{mordasini2016}. 
%\cite{mordasini2016,shibata2020,hands2022}.
Therefore, in general terms, the characteristics of WASP-193\,b, along with its exquisite suitability for atmospheric characterization, position it as a benchmark planet for gaining insights into the properties of the population of low-density planets.

\newpage

%%%%%%%%%%%%%% Main Table system properties
\begin{table*}
		{\renewcommand{\arraystretch}{1.5}
				\resizebox{1.\textwidth}{!}{% }
			\begin{tabular}{llcccc}
				\hline
				Parameter & Symbol & Value             & Value & Unit  \\
				          &        & (Circular, $e=0$) & (Eccentric, $e \neq 0$)  & \\
				\hline
				\textit{Deduced stellar parameters}  & \textbf{WASP-193}& \\
				\hline
				Mean density    &  $\rho_\star$   & $0.567_{-0.048}^{+0.051}$ &   $ 0.557_{-0.047}^{+0.051}$  & $\rho_\odot$        \\
				Stellar mass    &  $M_\star$       & $1.068 \pm 0.066$        &   $ 1.059_{-0.068}^{+0.067}$  & $M_\odot$            \\
				Stellar radius  &  $R_\star$       & $1.235 \pm 0.027$ &    $ 1.239\pm 0.028$  & $R_\odot$   \\
				Luminosity      & $L_\star$        & $1.87 _{-0.16}^{+0.18}$   &   $  1.88_{-0.16}^{+0.18}$ &  $L_\odot$    \\
				\hline
				\textit{Deduced planet parameters}  & \textbf{WASP-193 b}   \\
				\hline
				Scaled semi-major axis & $a/R_\star$   & $11.81 \pm 0.34$   &   $ 11.74 ^{+0.35}_{-0.34}$  & $R_\star$     \\
				Orbital semi-major axis  & $a$          &  $ 0.0678 \pm 0.0014$      &   $  0.0676 \pm 0.0015$      & au       \\
				Orbital inclination  &  $i_p$           & $ 88.51 _{-0.43}^{+0.57}$  &  $  88.49 _{-0.49}^{+0.78}$ & deg    \\
				Eccentricity  & $e$                     & $0$ (fixed)                  &   $  0.056^{+0.068}_{-0.040}$         &  -- \\
				Density   & $\rho_p$                    & $ 0.059^{+0.015}_{-0.014}$ &  $  0.059^{+0.015}_{-0.013}$ & ${\rm g~cm^{-3}}$            \\
				Surface gravity & $\log g_p$            & $ 2.23 _{-0.11}^{+0.09}$ &  $ 2.22 _{-0.11}^{+0.09}$  & cgs       \\
				Mass               & $M_p$              & $ 0.141^{+0.029}_{-0.030}$  &   $  0.139 \pm 0.029$      & $M_{\rm Jup}$           \\
				Radius          &  $R_p$                & $ 1.463^{+0.059}_{-0.057}$  &   $ 1.464^{+0.059}_{-0.057}$ &$R_{\rm Jup}$   \\
				Equilibrium temperature & $T_{\rm eq}$  & $ 1252 \pm 30$             &   $ 1254 \pm 31$ &  K           \\
				Irradiation  & $S_p$ & $5.58^{+0.57}_{-0.52} \times 10^5 $        &  $ 5.62^{+0.53}_{-0.57} \times 10^5 $ &$\text{W}  \text{m}^{-2}$  \\
				%Bayesian Information Criterion         &  BIC   & 2324    &  2318 &      -- \\
				\hline
		\end{tabular}}}
	\caption{The WASP-193 system parameters derived from our global MCMC analysis  (medians and $1\sigma$ limits of the marginalized posterior probability distributions).}
	\label{mcmcwasp193b}
\end{table*}

%%%
%%%
%%%
%%% METHODS
%%%
%%%
%%%

\clearpage

\begin{methods} 
\label{methods}

\subsection{WASP observations and data reduction}  
\textcolor{blue}{ }

\noindent
The Super-WASP (Wide Angle Search for Planets\cite{Pollacco2006}) project consists of data obtained at telescopes located at two sites:  Sutherland Station of the South African Astronomical Observatory (SAAO) and the Observatorio del Roque de los Muchachos on the island of La Palma in the Canary Islands. The field-of-view (FOV) of each instrument is 482~$\text{deg}^2$  with a pixel scale of 13.7''~pixel$^{-1}$. These instruments are capable of obtaining photometry with a precision better than 1\% for objects with  $V$-magnitudes between $ 7.0$ and $11.5$\cite{Pollacco2006}.
%%%%%%
The WASP  ground-based transit survey has discovered almost 200 planets, transiting bright nearby stars, mostly hot Jupiters plus minor fraction of short-period Saturn- and Neptune-mass objects. 
\noindent
Many WASP planets have proven to be among the most favorable targets for detailed atmospheric characterization with current facilities such as the \emph{HST}\cite{Meylan_2004PASP_HST}, \emph{VLT}\cite{Pepe_VLT}, \emph{Magellan}\cite{Dressler_2011_Magellan}, and \emph{JWST}\cite{Gardner_2006SSRv_jwst} (e.g., WASP-17\,b\cite{Sing2016}, WASP-101\,b\cite{Wakeford_2017_WASP-101b}, WASP-121\,b\cite{Sing_2019_WASP-121b} and WASP-127\,b\cite{Skaf_2020_WASP-127b}).
%%%%%%%

\noindent
The host-star WASP-193 was observed by the WASP-South survey in 2006 and 2012. The WASP data calibration (bias and dark subtraction, and flat-field division) was performed using a bespoke pipeline\cite{Cameron2006} developed in Fortran. Aperture photometry was performed on the final calibrated images, where the stars' fluxes were measured in three photometric apertures with radii of 2.5, 3.5, and 4.5 pixels, following the method described by ref.\cite{Christian2004,Cameron2006,Pollacco2006}. The WASP phase-folded light curve obtained for WASP-193 is presented in Supplementary Fig.~1. 
%%%%%%%
The search for transit events in the WASP photometry was conducted employing the Box-Least-Square (BLS) method as described in ref.\cite{Cameron2006}. The alert for this candidate triggered the ground-based follow-up campaign using both photometry and spectroscopy measurements to confirm the planetary nature of the detected signal.

\subsection{TRAPPIST-South photometry} 
\textcolor{blue}{ }

\noindent
TRAPPIST (TRAnsiting Planets and PlanetesImals Small Telescope) network is composed of the 0.6-m twin robotic telescopes TRAPPIST-South and TRAPPIST-North\cite{Jehin2011,Gillon2011,barkaoui2019}. For WASP-193, we used TRAPPIST-South located at the La Silla observatory in Chile. It is equipped with a thermo-electrically 2K$\times$2K FLI ProLine PL3041-BB CCD camera with a FOV of 22$^{\prime}$ $\times$ 22$^{\prime}$ and a pixel scale of 0.65$^{\prime\prime}$ per pixel. TRAPPIST-South observed a full transit of WASP-193\,b in the Sloan-$z'$ filter on 27 January 2017 with an exposure time of 20s,  three partial transits, one in the $I+z$ filter on 02 January 2015 with an exposure time of 10s, one in the $BB$ filter on 06 April 2015 with an exposure time of 8s, and a last one in the Johnson-$B$ filter on 08 June 2019 with an exposure time of 25s. Data calibration and differential photometry were performed using the \textit{PROSE} pipeline\cite{garcia2021}. The reduced light curves obtained for WASP-193\,b along with the best fitting model are presented in Fig.~\ref{Photometry_RVs} and Supplementary Table~1. 

\subsection{SPECULOOS-South photometry} 
\textcolor{blue}{ }

\noindent
The SPECULOOS (Search for habitable Planets EClipsing ULtra-cOOl Stars) robotic telescope network is composed of the SPECULOOS Southern Observatory\cite{Burdanov2017,Delrez2018,Jehin2018Msngr} (SSO), with four 1-m telescopes at ESO Paranal Observatory in Chile and two 1-m telescopes in the Northern hemisphere: the SPECULOOS Northern Observatory\cite{burdanov2022} (SNO) at Teide Observatory in Spain and SAINT-EX\cite{demory2020} (Search And characterIsatioN of Transiting EXoplanets) at San Pedro Mártir Observatory in Mexico. These facilities are identical Ritchey-Chretien telescopes equipped with ANDOR iKon-L BEX2-DD cameras and 2048$\times$2088 e2v CCD detectors, with a FOV of 12$^{\prime}$ $\times$ 12$^{\prime}$ and a pixel scale of 0.35$^{\prime\prime}$ per pixel. We observed a partial transit of WASP-193\,b with one of the four telescopes of SSO on 08 June 2019 in the Johnson-$B$ filter with an exposure time of 25s. Data calibration and differential photometry were performed using the \textit{PROSE} pipeline\cite{garcia2021}. The reduced light curve obtained jointly with the best-fitting model is presented in Fig.~\ref{Photometry_RVs} and Supplementary Table~1.  

\subsection{\emph{TESS} photometry} 
\textcolor{blue}{ }

\noindent
The \emph{TESS} (Transiting Exoplanet Survey Satellite\cite{Ricker2015}) mission was launched by NASA in April 2018. Its main goal is to detect and characterise transiting exoplanets smaller than Neptune orbiting nearby bright stars. \emph{TESS}'s combined FOV is $24^\circ\times 96^\circ$, and one pixel is $\sim21$'' on the sky.  WASP-193 was observed by \emph{TESS} with a 30-min cadence in the ninth sector of its primary mission (from 28 February to 26 March 2019), with a 10-min cadence during its extended mission in sector 36 (from 07 March to 02 April 2021) and with 2-min cadence in sector 63 (from 10 March 2023 to 06 April 2023).
We retrieved the light curves produced by the TESS Science Processing Operations Center pipeline\cite{Stumpe_2012PASP,Smith_2012PASP,Stumpe_2014} (PDC-SAP) corresponding to the 2-min and 10-min cadences from the Mikulski Archive for Space Telescope.  These light curves are corrected for instrument systematics and crowding effects. \emph{TESS} light curves corresponding to the 2-min and 10-min cadences for WASP-193\,b and the best fitting model are presented in Fig.~\ref{Photometry_RVs} and Supplementary Table~1.

%The data calibration was done by the SPOC pipeline (\emph{TESS} Science Processing Operations Center) as described by ref.~\cite{Jenkins2016SPIE}. 
%Photometric reduction was then performed  with the \href{https://photutils.readthedocs.io/en/stable/#user-documentation}{\it Python/Photutils} package. The resulting light curve of WASP-193 is presented in  \autoref{lcs_TS_SSO}.

\subsection{Spectroscopic and radial-velocity measurements} \label{spectro_analysis}
\textcolor{blue}{ }

\noindent
In this paper, we used RV measurements obtained by CORALIE (mounted on the 1.2-m Swiss Euler telescope located at ESO La Silla in Chile\cite{Queloz2000}) and HARPS (mounted on the 3.6-m telescope at ESO La Silla in Chile\cite{Mayor2003Msngr}) spectrographs. We obtained 12 (between June 2015 and June 2018) and 11 (between February and July 2019) spectroscopic measurements of WASP-193 with the CORALIE and HARPS spectrographs, respectively. We applied the cross-correlation method\cite{Baranne1996} on the observed spectra of WASP-193 to extract the radial velocity measurements.

\noindent 
The RV measurements of WASP-193 are in phase with the ephemerides from the WASP-South photometric data, but the planet was not independently detected in the RV data.  
The offset between RV data sets is taken into account by modeling the systematic velocity for each instrument during the global analysis. The marginally strongest signal in the generalised Lomb-Scargle periodogram\cite{Zechmeister_2009AA496.577Z} coincided with the transit period of 6.25 days, though the significance was below 10\% FAP. 
The RV measurements obtained with CORALIE and HARPS are presented in Supplementary Table~2, and the phase-folded RV curve of WASP-193 is presented in Fig.~\ref{Photometry_RVs}.
The corresponding RV based on the fitting linear slope, enables us to discard the blended eclipsing binary (BEB) scenario and to keep the planetary companion hypothesis.  The correlation between the bisector-spans and RVs is displayed in Fig.~\ref{Photometry_RVs} panel $d$, which was consistent with zero, i.e., there is no significant correlation between these two parameters. Moreover, we calculated the Pearson $r$ coefficient to measure the correlation between these two parameters, and we obtained $r_{\rm Harps} = 0.07$ for HARPS and $r_{\rm Coralie} = 0.21$ for CORALIE. Therefore, the Pearson $r$ coefficient confirmed no significant correlation between the bisector-spans and RVs, supporting the planetary nature of WASP-193\,b.

 %The deduced RV offset between the two instruments is $2 \pm 9.15$~m/s  from the global fit. When fixing the offset between CORALIE and HARPS to 2~m/s, 

\noindent
 We used co-added spectra obtained with the  HARPS spectrograph to produce a single spectrum with S/N of 75, which allowed us to constrain the stellar atmospheric parameters of WASP-193 better using the method described by ref.\cite{Doyle2013}. The metallicity, $[Fe/H]$, was computed from the Fe lines, the effective temperature, $T_{\rm eff}$, was computed from the H$\alpha$ line, the surface gravity, $\log g$, was estimated from the Mg~{\rm \sc I}b and Na~{\rm \sc I}D lines,  the activity index, $\log (R'_{hk})$, was constrained from the Ca~{\sc II} H+K lines and the rotational velocity, $V\sin(i)$, was measured by assuming a macroturbulence value of $v_{\rm macro} = 4.45$~km~s$^{-1}$, which was extracted from the calibration formula\cite{Doyle_2014}, and a microturbulence value of $v_{\rm micro} = 1.17$~km~s$^{-1}$, which was extracted from the calibration formula\cite{Bruntt_2010}. With these values, it was found that WASP-193 is a Sun-like star with $T_{\rm eff} = 6076 \pm 120$~K, $\log g_\star = 4.1 \pm 0.1$~dex, $[Fe/H] = -0.06 \pm 0.09$~dex, $V\sin (i) = 4.3 \pm 0.8$~km~s$^{-1}$,  and $\log (R'_{hk}) = 5.30 \pm 0.07$~[dex]. Using the {\it MKCLASS} program\cite{Gray_2014AJ_MKCLASS},  we got a spectral-type of F9 from the observed HARPS spectrum. This is consistent with that estimated from the stellar atmospheric parameters and  $B-V$ color index.

\subsection{Stellar parameters from the {\it ExoFASTv2} analysis} \label{exofastv2}
\textcolor{blue}{ }

\noindent
As an independent check of the derived stellar parameters for WASP-193, we performed an analysis using the exoplanet fitting suite, {\it EXOFASTv2}\cite{Eastman_2013PASP,Eastman_2019PASP}.
{\it EXOFASTv2} uses a differential evolution MCMC to globally and simultaneously model the star and planet. The built-in Gelman--Rubin statistic\cite{Gelman1992,Gelman2003} is used to check the convergence of the chains. A full description of {\it EXOFASTv2} was given by  ref.\cite{Eastman_2019PASP}. Within the fit, the host star's parameters are determined using the SEDs (see Supplementary Fig.~2) constructed from our broadband photometry, MESA Isochrones and Stellar Tracks (MIST) evolutionary models\cite{Choi:2016, Dotter:2016}.
For the SED fits performed within  {\it EXOFASTv2}, we used photometry from APASS DR9 BV\cite{Henden2016}, 2MASS HK\cite{Cutri:2003}, and ALL-WISE W1, W2 and W3\cite{Wright:2010}, which are presented in Supplementary Table~3. 

\noindent
We applied Gaussian priors to the parameters $T_{\rm eff}$, $\log g_\star$, and $[Fe/H]$ using the obtained values from the spectroscopic analysis.  We also enforced Gaussian priors on the Gaia~DR3\cite{Gaia_Collaboration_2020_DR3} parallax (adding 82~$\mu$as to the reported value and adding 33~$\mu$as in quadrature to the error, following the recommendation of ref.\cite{Stassun:2018}). We adopted an upper limit on the extinction of $A_V=0.20$ from ref.\cite{Schlafly:2011}. The {\it EXOFASTv2} fit results are presented in Supplementary Table~3.

% using their web interface\footnote{{ }\url{https://irsa.ipac.caltech.edu/applications/DUST/}}

\subsection{Stellar parameters using {\it CLES}} \label{cles}
\textcolor{blue}{ }

\noindent
We used our Li\`ege stellar evolution code, {\it CLES}, to determine the age and mass of the host star. All the details concerning the input physics (treatment of convection, surface boundary conditions, nuclear reaction rates, equation of state, opacities, and diffusion) can be found in Sect.~2.3 of ref.\cite{Salmon_2021_Stellar_evol}. We adopted the solar mixture of ref.\cite{Asplund_2009ARA&A}. As inputs, we used the spectroscopic $T_{\rm eff}$ and $[Fe/H]$, as well as the stellar radius $R_\star$ obtained from SED-fitting using {\it ExoFASTv2}. We computed stellar evolutionary tracks and selected the best-fit one according to the input parameters following the Levenberg-Marquadt minimisation scheme, as explained in ref.\cite{Salmon_2021_Stellar_evol}. We obtained $M_\star=1.102\pm0.070 M_{\odot}$ and $t_\star=4.4\pm1.9$ Gyr (see Supplementary Table~3). 

% \footnote{Very similar results are obtained adopting instead the luminosity $L_\star$ obtained from bolometric flux (again by SED-fitting) and distance, or the stellar density derived from transit light curves.}

\subsection{Modelling of the spectroscopic and photometric data} \label{Data_analys}
\text{ }

\noindent
We combined the spectroscopic and photometric observations of WASP-193 to determine the physical parameters of the planetary system. The stellar atmospheric parameters, $T_{\rm eff}$, $\log g_\star$ and $[Fe/H]$ from our spectroscopic analysis, were coupled with the stellar radius, $R_\star$, obtained using {\it EXOFASTv2} and the mass, $M_\star$, obtained using {\it CLES} to infer the stellar density, $\rho_\star$, luminosity, $L_\star$, and other physical parameters.

\noindent
We used a Markov-chain Monte Carlo (MCMC) algorithm\cite{Gillon2012} to sample the posterior probability distributions of the system's parameters, from which we extracted the median values and their $1\sigma$-uncertainties.
All photometric measurements contain additional faint ($G_{\rm mag}\sim 16$) neighbour star at 4.2" (see Supplementary Fig.~3). The resulting dilution was found to be $<5\%$, which is included in our global analysis.

\noindent
We modelled the RV curves using a 2-body Keplerian model\cite{Murray2010}, and the transit light curves using the Mandel \& Agol\cite{Mandel2002} model multiplied by a baseline to correct for several external effects related to time, full-width half-maximum (FWHM), airmass, star position in the CCD, $x$ and $y$, and background (see Supplementary Table~1). 
The baseline model for each transit light curve was selected based on minimizing the Bayesian information criterion (BIC)\cite{schwarz1978} using the formula:

\begin{equation}
BIC = \chi^2 + k\log (N),
\end{equation}

\noindent
 where $k$ is the number of free parameters, $N$ is the number of data points, and $ \chi^2 = -2\log(L)$, $L$ is the maximized value of the model's likelihood function. The light curve observed on 06 April 2015 with TRAPPIST-South contains a meridian flip (i.e., a 180$^\circ$ rotation that the German equatorial mount has to undergo when the target star passes the local meridian) at BJD $ = 2457119.619198$ (see details in Supplementary Table~1),  that we modeled as a flux offset in our MCMC analysis.

\noindent
The jump parameters in our MCMC analysis (i.e., parameters  randomly perturbed at each step of the MCMC, see Supplementary Table~4) were:
\begin{itemize}
	\item[$\bullet$]  $T_0$: the transit timing;
	\item[$\bullet$] $W$: the total duration of the transit (i.e. duration between ingress and egress);
	\item[$\bullet$] $dF = R_p^2/R_\star^2$: the transit depth, where $R_p$ and $R_\star$ are the planetary and stellar radii, respectively;
	\item[$\bullet$] $b' = a \cos(i) / R_\star$: the impact parameter, where $i$ is the orbital inclination and $a$ is the orbital semi-major axis;
	\item[$\bullet$] $P$: the orbital period;
	\item[$\bullet$]  $K_2 = K~P^{1/3}\sqrt{1-e^2}$, where $K$ is the semi-amplitude of the RV and $e$ is the orbital eccentricity; and
	\item[$\bullet$]  $\sqrt{e}\cos(w)$ and $\sqrt{e}\sin(w)$, where $w$ is the argument of periastron.
\end{itemize}

\noindent
The stellar parameters $T_{\rm eff}$, $[Fe/H]$, and $M_\star$ were also jump parameters in our analysis. For $T_{\rm eff}$, $[Fe/H]$, and $\log g_\star$, we applied Gaussian priors based on the results of our spectroscopic analysis. 

\noindent
We assumed a quadratic limb-darkening (LD) law to take into account the impact of limb-darkening on our transit light curves. For each filter, the quadratic LD coefficients, $u_1$ and $u_2$, were interpolated from the tables of ref.\cite{Claret2011} using the $T_{eff}$, $\log g_\star$ and $[Fe/H]$, obtained from our spectroscopic analysis. For the non-standard $I+z$ and $BB$ filters, we took the averages of the values for the standard filters $Ic$ and Sloan-$z'$ for the $I+z$ filter, and $Ic$ and Johnson-$R$ for the $BB$ filter. The LD coefficient values of $u_1$ and $u_2$ for each filter are presented in Supplementary Table~5.

\noindent
We performed two independent MCMC analyses: one assuming a circular orbit (model 1) and the other assuming an eccentric orbit (model 2). For each model, we calculated the BIC value, which was then used to compute the Bayes Factor (BF) using the relationship described by ref.\cite{wangenmarkers2007}. The Bayes Factor (BF$_{12}$) was determined as BF$_{12}$ = $e^{(BIC_{model2} - BIC_{model1})/2}$. Our analysis yielded a BF$_{12}$ value of 20.1, indicating strong evidence in favor of the eccentric scenario\cite{wangenmarkers2007}.

\noindent
For each transit light curve, a preliminary MCMC analysis composed of one Markov chain of 100\,000 steps was performed to estimate the correction factor ($CF$) to rescale the photometric errors by $CF = \beta_{\rm red} \times \beta_{\rm white}$, where $\beta_{\rm red}$ is the red noise and $\beta_{\rm white}$ is the white noise, as described by ref.\cite{Gillon2012}. For the RVs, a $jitter$ noise (i.e., the quadratic difference between the mean error of the measurements and the standard deviation of the best-fitting residuals) of 22.3~m~s$^{-1}$ for CORALIE and 4.1~m~s$^{-1}$ for HARPS was added quadratically to the RV error bars. Then, we conducted a global MCMC analysis composed of five chains of 100\,000 steps. The convergence of each chain was checked using the statistical test presented by Gelman-Rubin\cite{Gelman1992}. The transit depths derived for different photometric bands are consistent at the 1$\sigma$ level, highlighting the non-chromatic dependence and supporting our planetary interpretation (see Supplementary Table~4). The deduced parameters of the WASP-193 system are presented in Table~\ref{mcmcwasp193b}. The stellar and planetary physical parameters posterior probability distributions are shown in Supplementary Fig.~4 and Supplementary Fig.~5, respectively. 

\subsection{Atmospheric model and injection-retrieval test}

\text{ }

\noindent
The atmospheric model and the injection-retrieval test performed to assess WASP-193~b's suitability for atmospheric exploration with JWST uses the \texttt{TIERRA} retrieval framework\cite{Niraula_2022NatAs}. We model a synthetic WASP-193~b atmosphere using abundances consistent with WASP-39\,b (volume mixing ratio of CO:10$^{-5}$, CO$_2$: 10$^{-4}$, H$_{\rm 2 }$
O: 10$^{-3}$, SO$_2$:10$^{-5}$). We tested four different instrumental settings (NIRSpec Grism, NIRISS SOSS, NIRCAM F322W2, and NIRSpec G395H) and found that while all lead constraints on the planetary properties, \emph{JWST/NIRSpec} Prism provides the most comprehensive constraints owing to its more extensive wavelength coverage. Fig.~\ref{fig:Synth_Spectra} presents the synthetic data, the best fit, and the retrieved planetary parameters.

\end{methods}

%%%
%%%
%%%
%%% ADDENDUM
%%%
%%%
%%%

\begin{addendum}
\item[Data Availability] {The TRAPPIST-South and SPECULOOS-South data analysed are available at CDS via anonymous ftp to  in \url{http://cdsarc.u-strasbg.fr/} (130.79.128.5).
The HARPS and CORALIE data analysed are available in the European Southern Observatory archive (ESO; \url{http://archive.eso.org/cms.html})
The TESS data analysed are available in the Mikulski Archive for Space Telescopes (MAST; \url{https://archive.stsci.edu/}).
}
% \item[Code Availability] 
% The code \textit{Prose} used to reduce the TRAPPIST-South and SPECULOOS-South data
% The code \textit{MCMC}\cite{Gillon2012} used to analyse the transit light curves.
% Additioanl packages used in this study are:
% \textit{Matplotlib}\cite{Matplotlib}, \textit{Numpy}\cite{Numpy},
% \textit{Astropy}\cite{Astropy},
% \textit{SciPy}\cite{Scipy},
% \textit{Lightkurve}\cite{Lightkurve_2018ascl.soft12013L},
% \textit{Wotan}\cite{Hippke2019AJ_wotan},
% \textit{Photutils}\cite{larry_bradley_2020_4044744},
% \textit{EXOFATSv2}\cite{Eastman_2019PASP},
% \textit{petitRADTRANS}\cite{Molliere_2019_petitRADTRANS}.
% %\textit{BAGEMASS}\cite{Maxted_2015AandA}.
\item[Code Availability] 
The \textit{Prose} and \textit{MCMC} codes used for this paper are publicly available.

\end{addendum}

\begin{addendum}
% \newpage

\item[Acknowledgments]
WASP-South is hosted by the South African Astronomical Observatory and we are grateful for their ongoing support and assistance. Funding for WASP comes from consortium universities and from the UK's Science and Technology Facilities Council. The research leading to these results has received funding from the European Research Council (ERC) under the FP/2007–2013 ERC grant agreement no. 336480, and under the H2020 ERC grant agreement no. 679030; and from an Actions de Recherche Concertée (ARC) grant, financed by the Wallonia-Brussels Federation. The Euler Swiss telescope by the Swiss National Science Foundation (SNF). This work has been carried out in part within the framework of the NCCR PlanetS supported by the Swiss National Science Foundation.
%% ESO 
This study is based on observations collected at the European Southern Observatory under ESO programme 0102.C-0414, PI: Nielsen. 
%%%% TRAPPIST-South
TRAPPIST-South is funded by the Belgian National Fund for Scientific Research (F.R.S.-FNRS) under grant PDR T.0120.21, with the participation of the Swiss National Science Fundation (SNF). MG is F.R.S.-FNRS Research Director and EJ is F.R.S.-FNRS Senior Research Associate. V.V.G. is an F.R.S.-FNRS Research Associate.
%% KB and MT
The postdoctoral fellowship of KB is funded by F.R.S.-FNRS grant T.0109.20 and by the Francqui Foundation.
This publication benefits from the support of the French Community of Belgium in the context of the FRIA Doctoral Grant awarded to MT.
%%%% SPECULOOS
The ULiege's contribution to SPECULOOS has received funding from the European Research Council under the European Union's Seventh Framework Programme (FP/2007-2013) (grant Agreement n$^\circ$ 336480/SPECULOOS), from the Balzan Prize Foundation, from the Belgian Scientific Research Foundation (F.R.S.-FNRS; grant n$^\circ$ T.0109.20), from the University of Liege, and from the ARC grant for Concerted Research Actions financed by the Wallonia-Brussels Federation. 
This work is supported by a grant from the Simons Foundation (PI Queloz, grant number 327127).
J.d.W. and MIT gratefully acknowledge financial support from the Heising-Simons Foundation, Dr. and Mrs. Colin Masson and Dr. Peter A. Gilman for Artemis, the first telescope of the SPECULOOS network situated in Tenerife, Spain.
This work is supported by the Swiss National Science Foundation (PP00P2-163967, PP00P2-190080 and the National Centre for Competence in Research PlanetS).
This work has received fund from the European Research Council (ERC)
under the European Union's Horizon 2020 research and innovation programme (grant agreement n$^\circ$ 803193/BEBOP), from the MERAC foundation, and from the Science and Technology Facilities Council (STFC; grant n$^\circ$ ST/S00193X/1).
C.D. was supported by the Swiss National Science Foundation (SNSF) under grant PZ00P2\_174028.
M.L. acknowledges support of the Swiss National Science Foundation under grant number PCEFP2\_194576.
E. D acknowledges support from the innovation and research Horizon 2020 program in the context of the  Marie Sklodowska-Curie subvention 945298.
F.J.P acknowledges financial support from the grant CEX2021-001131-S funded by MCIN/AEI/ 10.13039/501100011033 and through projects PID2019-109522GB-C52 and PID2022-137241NB-C43. 
\item[Author Contributions Statement]
K.B. led the project and performed and interpreted the global analyses with support from  F.J.P. and M.G.. F.J.P. led the interpretation of the results and writing of the paper.
P.N. and J.d.W. performed the assessment of atmospheric-characterization suitability.
C.H., O.T., D.R.A., S.U., and R.G.W. performed the WASP-South observation and data reduction.
C.D., R.H., and S.M. performed planet interior models and interpretation of the results.
B.S., V.V.G., P.F.L.M., A.S., and M.G. performed spectroscopic data, stellar evolutionary models, and SED analysis.
L.D.N., F.B., and M.L. performed radial velocity measurements.
E.J., E.D., C.A.M., and P.P.P. performed photometric follow-ups using TRAPPIST-South and SPECULOOS-South facilities and data reduction using Prose.
All co-authors read and commented on the manuscript and helped with its revision.

\item[Competing Interests Statement] The authors declare no competing interests.

\item[Correspondence] Correspondence and requests for materials should be addressed to K.B. (khalid.barkaoui@uliege.be) and F.J.P. (pozuelos@iaa.es).

\end{addendum}

%%%
%%%
%%%
%%% MAIN FIGURES
%%% Letters should have no more than four display items (figures and/or tables).
%%%
%%%

\pagebreak
%\clearpage

\pagebreak

\pagebreak

\supplement

%%%%%%%%%%%%%%%%%%%%%%%%%
%%%%%%%%%%%%%%%%%%%%%%
%% Suplement info
%%%%%%%%%%%%%%%%%%%%%%%
%%%%%%%%%%%%%%%%%%%%%%%%%%

%%%% TABLES %%%%%

%%%%%% Ground-based observations table
\begin{table*}[!h]
	\centering
	{\renewcommand{\arraystretch}{1.3}
		\resizebox{1\textwidth}{!}{% }
			\begin{tabular}{lcccccc|cccccc}
				\hline
				\multicolumn{7}{c|}{Observation parameters} & \multicolumn{5}{c}{MCMC analysis} \\
				\hline
				Telescope & Date &  Filter  & $N_p$  & $ExpT$ (s) & Duration & $FWHM$ & Baseline  & Residual RMS & $\beta_w$ & $\beta_r$ & $CF$ \\
				&        &          &      &   & (min) & (pixel) & function & &&& \\
				\hline\hline
				TRAPPIST-S & 02 Jan 2015 &  $I+z$        & 638 &   10 & 224 & 2.90 &    $f(t^1,f^1)$ &  0.0043 & 1.09 & 2.27 & 2.48  \\
				 TRAPPIST-S   & 06 Apr 2015 &  $BB$       & 724 &    8  & 230 & 2.67 & $f(a^1,O)$ & 0.0037 & 0.85  & 1.65 & 1.40\\
			 TRAPPIST-S &  27 Jan 2017 &  $z'$ & 602 &  20  & 297 & 3.28 & $f(t^1,b^1)$ & 0.0042 & 1.00 & 1.34  & 1.34\\
				TRAPPIST-S   &  08 Jun 2019 &  $B$            & 292 & 25 & 176   & 3.32 & $f(t^1,b^1)$ & 0.0042 & 0.74 & 1.60 & 1.18 \\
				SPECULOOS-S  & 08 Jun 2019 & $B$             & 484 & 15 & 280   & 5.02 & $f(t^1,a^1)$ & 0.0034 & 0.63 & 1.93 & 1.20 \\
             TESS &  07/03 -- 02/04/2021  &   TESS & - & 600 &  - & - & - &  0.0012 &  0.94 &  1.00 &  0.94  \\
             TESS &  10/03 -- 06/04/2023  &   TESS & - & 120 &  - & - & - &  0.0025 &  0.97 &  1.02 &  0.99  \\
				\hline
	\end{tabular}}}
	\caption*{\textbf{Supplementary Table 1.} Observational and MCMC analysis parameters. {\it Observation parameters}:  date of observation,  telescope,  filter(s) used,   number of images,  exposure time(s), observation duration and FWHM of the point-spread function. {\it MCMC analysis parameters:}  selected baseline-function,  standard deviation (RMS) of the best-fitting residuals,  deduced values of $\beta_{\rm white}$, $\beta_{\rm red}$ and the coefficient correction  $CF =  \beta_{\rm white} \times \beta_{red}$. For the baseline-function, $f(\varepsilon^n)$, denotes, respectively,  the $n-$order polynomial function of the airmass ($\varepsilon=a$), FWHM ($\varepsilon=f$), background sky ($\varepsilon = b$), and the time ($\varepsilon = t$). The symbol $O$ denotes an offset fixed at the time of the meridian flip.}
	\label{baseline_model}
\end{table*}

\pagebreak

%%%%%% Rv data
\begin{table*}[!ht]
\centering
	{\renewcommand{\arraystretch}{1.1}
		\begin{tabular}{lccc}
			\hline
			BJD-TDB - 2.450.000     & RV [Km/s] & $\sigma_{RV}$ [Km/s] & Instrument      \\
			\hline\hline
			7189.495005 &	-3.20480 &	0.03043	 & Euler-1.2m/CORALIE \\
			7192.541007 & 	-3.21921 &	0.07803	 & Euler-1.2m/CORALIE\\
			7370.835932 &	-3.21178 &	0.03220 & Euler-1.2m/CORALIE\\
			7399.808035 &	-3.13149 &	0.04401 & Euler-1.2m/CORALIE\\
			7407.794853 &	-3.20625 &	0.02373 & Euler-1.2m/CORALIE\\
			7433.698835	& -3.18020 &	0.02095 & Euler-1.2m/CORALIE\\
			7748.849031	& -3.24191 &	0.05906 & Euler-1.2m/CORALIE\\
			7750.849267	& -3.22754 &	0.03067 & Euler-1.2m/CORALIE\\
			7752.804569	& -3.18884	& 0.05035 & Euler-1.2m/CORALIE\\
			7917.511102	& -3.16284 &	0.02673 & Euler-1.2m/CORALIE\\
			8277.509694 &	-3.15936 &	0.03543 & Euler-1.2m/CORALIE\\
			8298.463390 &	-3.16139 &	0.03678 & Euler-1.2m/CORALIE\\
			\hline
			8520.833792 &	-3.20823 &	0.01026 & ESO-3.6m/HARPS \\
			8537.867181 &	-3.17957 &	0.00695 & ESO-3.6m/HARPS \\
			8539.862413	& -3.17219	& 0.00829 & ESO-3.6m/HARPS \\
			8540.782640	& -3.18160	& 0.00684 & ESO-3.6m/HARPS \\
			8541.860510	& -3.16300	& 0.00551 & ESO-3.6m/HARPS \\
			8542.824208	& -3.16043 &	0.00537 & ESO-3.6m/HARPS \\
			8543.807974	& -3.18210 &	0.00820 & ESO-3.6m/HARPS \\
			8660.500666	& -3.14632 &	0.01434 & ESO-3.6m/HARPS \\
			8661.541276	& -3.16883	& 0.01107 & ESO-3.6m/HARPS \\
			8670.484865	& -3.18669 &	0.00619 & ESO-3.6m/HARPS \\
			8684.509150	& -3.16585	& 0.01232 & ESO-3.6m/HARPS \\
			\hline
		\end{tabular}}
		\caption*{\textbf{Supplementary Table 2.} Radial-velocities measurements for WASP-193 obtained from Euler-1.2m/CORALIE and ESO-3.6m/HARPS spectrographs.}
		\label{table_rv_wasp193}
\end{table*}

\pagebreak

%%%%%% Stellar properties table
\begin{table*}[!ht]
\centering
	{\renewcommand{\arraystretch}{0.9}
       \resizebox{0.995\textwidth}{!}{% }
		\begin{tabular}{lcc}
			\hline
			\hline
			\multicolumn{3}{c}{  Star information}   \\
			\hline
			\hline
			Parameter & WASP-193 & Source   \\
			\hline
			Identifying information: & & \\
			WASP\,ID  & 1SWASPJ105723.88-295949.6 &\\
			GAIA\,DR3\,ID  & 5453063823882876032 & Ref.~\cite{Gaia_Collaboration_2020_DR3}\\
			  TESS\,ID   &   TIC 49043968   & Ref.~\cite{Stassun_2018AJ_TIC} \\
			2MASS\,ID & 10572385-2959497 & Ref.~\cite{Cutri:2003}\\
			RA [J2000]     &   $ 10^h 57^m 23.85^s  $  & Ref.~\cite{Gaia_Collaboration_2020_DR3}   \\
			Dec [J2000]   &  $ -29^\circ 59' 49.66'' $  & Ref.~\cite{Gaia_Collaboration_2020_DR3} \\
			\hline
			Parallax  and distance: & \\
			Plx [$mas$] &   2.648 $\pm$ 0.015  & Ref.~\cite{Gaia_Collaboration_2020_DR3} \\
			Distance [pc] &    377.72 $\pm$ 2.17  & Ref.~\cite{Gaia_Collaboration_2020_DR3} \\
			\hline
			Photometric properties: & \\
			$V_{\rm mag}$ [APASS]   &    12.19 $\pm$ 0.09 & Ref.~\cite{Henden2016}     \\
			$B_{\rm mag}$ [APASS]    &    12.72 $\pm$ 0.03 & Ref.~\cite{Henden2016}\\
            $G_{\rm mag}$ [Gaia\_DR3]   &     12.033 $\pm$ 0.003    & Ref.~\cite{Gaia_Collaboration_2020_DR3} \\
			$J_{\rm mag}$ [2MASS]   &     10.95   & Ref.~\cite{Cutri:2003}\\
			$H_{\rm mag}$ [2MASS]   &     10.81$\pm$ 0.03 & Ref.~\cite{Cutri:2003} \\
			$K_{\rm mag}$ [2MASS]   &     10.75 $\pm$ 0.04 & Ref.~\cite{Cutri:2003} \\		
			TESS$_{\rm mag}$    &     11.63 $\pm$ 0.01    & Ref.~\cite{Stassun_2018AJ_TIC} \\
			$W1_{\rm mag}$ [WISE] &  10.64 $\pm$ 0.03 & Ref.~\cite{Wright:2010}\\
			$W2_{\rm mag}$ [WISE] &  10.68 $ \pm$ 0.03  & Ref.~\cite{Wright:2010}\\
			$W3_{\rm mag}$ [WISE] &  10.49 $\pm$ 0.09  & Ref.~\cite{Wright:2010}\\
			\hline
			\multicolumn{2}{l}{Stellar parameters from spectroscopic analysis}  \\
			$T_{\rm eff}$ [K]  & 6076 $\pm$ 120   & See Methods \\
			$\log g_\star$ [dex]  &  4.1 $\pm$ 0.1  & See Methods \\
			$\log (R'_{hk})$ [dex] &    $5.30 \pm 0.07 $ & See Methods \\
			$[Fe/H]$ [dex] & -0.06 $\pm$ 0.09  & See Methods\\
			$V\sin (i)$ [km/s] &  4.3 $\pm$ 0.8  & See Methods\\
			$v_{\rm macro}$ [km/s] & 4.45  & See Methods\\
			$v_{\rm micro}$ [km/s] & 1.17 & See Methods\\
			Spectral type   & F9  & See Methods\\
            \hline
			\multicolumn{2}{l}{Stellar parameters from SED and {\it CLES} analysis}  \\
            $T_{\rm eff} [K]$ &    $ 6080^{+90}_{-98}$ & EXOFASTv2 (See Methods) \\
            $[Fe/H]$      &  $-0.193 \pm 0.086$ & EXOFASTv2 (See Methods) \\
            $M_\star [M_\odot]$ &  $  1.120 \pm 0.051$ & EXOFASTv2 (See Methods) \\
			  $M_\star [M_\odot]$ & $ 1.102\pm 0.070$ &  {\it CLES} (See Methods) \\
            $R_\star [R_\odot]$   & $  1.225^{+0.032}_{-0.029}$ & EXOFASTv2 (See Methods) \\
            $L_\star [L_\odot]$   & $1.65 ^{+0.13}_{-0.12}$& EXOFASTv2 (See Methods) \\
            $Age$ [Gyr]           &  $  4.4 \pm 1.9$ & {\it CLES}(See Methods) \\
            $Age$ [Gyr]           & $6.6 \pm 2.4$ & EXOFASTv2 (See Methods) \\
			\hline
	\end{tabular} }}
	%\caption{Stellar parameters for WASP-193.}
	\caption*{\textbf{Supplementary Table 3.} Stellar properties of WASP-193 derived from astrometry, photometry, spectroscopy, SED and {\it CLES} analysis.
	\label{stellarpar}}
	%\tablenotetext{test}
\end{table*}

%%%%%% Jump parameters during the MCMC fitting

\begin{table*}[!ht]
		{\renewcommand{\arraystretch}{1.5}
				\resizebox{1.\textwidth}{!}{% }
			\begin{tabular}{llcccc}
				\hline
				Parameter & Symbol & Value             & Value & Unit  \\
				          &        & (Circular, $e=0$) & (Eccentric, $e \neq 0$)  & \\
				\hline
				\textit{Jump parameters}      &  \\
				\hline
				Planet/star area ratio & $(R_p/R_\star)^2_{Ic+z'}$     & $1.42 \pm 0.08$          &  $ 1.41 \pm 0.08$   & \%      \\   
				                       & $(R_p/R_\star)^2_{{Sloan}-z'}$     & $1.30 \pm 0.09$          &  $ 1.31 \pm 0.09$   & \%      \\
				                       & $(R_p/R_\star)^2_{{\rm Johson}-B}$     & $1.44 \pm 0.18$          &  $ 1.45 \pm 0.18$   & \%      \\
				                       & $(R_p/R_\star)^2_{\rm Blue - Blocking}$     & $1.26 \pm 0.17$          &  $ 1.28 \pm 0.16$   & \%      \\
				                       & $(R_p/R_\star)^2_{TESS}$     & $1.24 \pm 0.11$          &  $ 1.24 \pm 0.11$   & \%      \\
				Impact parameter & $b' = a\cos i_p / R_\star$  & $0.306 _{-0.110}^{+0.077}$ &  $  0.309 _{-0.160}^{+0.094}$  & $R_\star$       \\
				Transit duration &  $W$                        & $0.1812 \pm 0.0018$        &  $ 0.1812 \pm 0.0022$ & days \\
				Transit-timing & $T_0$                         & $7781.66563 \pm 0.00055$   &   $ 7781.66567 \pm 0.00056$ & BJD$_{TDB}$ - 2450000     \\
				Orbital period  & $P$                          & $ 6.2463345 \pm 0.0000003$ &   $  6.2463345 \pm 0.0000003$ & days     \\ 
				RV semi-amplitude  & $K$                       & $14.9 \pm 3.1$             &   $ 14.8 \pm 3.0$  & m~s$^{-1}$        \\
				\hline
		\end{tabular}} }
	\caption*{\textbf{Supplementary Table 4.} Jump parameters used in our global MCMC analysis.}
	\label{Jump_Param}
\end{table*}

%%%%%% Priors for the global fitting table

\begin{table*}[!ht]
\centering
	\begin{center}		 
		{\renewcommand{\arraystretch}{1}
			%\resizebox{0.49\textwidth}{!}{% }
			\begin{tabular}{lcccc}
				\hline
				Parameter         &  Value &  Prior &       \\
				\hline\hline
				Effective temperature $T_{\rm eff}$ [K] &  $6078 \pm 120$   & $N(6080, 120^2)$  \\ 
				Surface gravity   $\log g_\star$ [dex]  & $4.284 \pm 0.033$ &  $N(4.1, 0.1^2)$ \\
				Rotational velocity $V\sin(i)$ [km/s]   & $4.29 \pm 0.81$ &  $N(4.3,0.8^2)$ \\
				Metallicity $[Fe/H]$ [dex]              & $-0.060 \pm 0.088$ &  $N(-0.06,0.09^2)$ \\
				Stellar mass $M_\star$ [$M_\odot$]      & $1.059 \pm 0.067$ &  $N(1.102,0.070^2)$ \\
                Stellar radius $R_\star$ [$R_\odot$]      & $1.235 \pm 0.027$ &  $N(1.225,0.032^2)$ \\
				Quadratic LD $u_{1, {\rm Sloan-}z'}$    & $0.207 \pm 0.013$ &  $N(0.207, 0.013^2)$    \\
				Quadratic LD $u_{2, {\rm Sloan-}z'}$    & $9.288 \pm 0.065$ &  $N(0.288,0.007^2)$ \\
				Quadratic LD $u_{1, I+z'}$              & $0.219 \pm 0.014$ &  $N(0.224,0.014^2)$  \\
				Quadratic LD $u_{2, I+z'}$              & $0.290 \pm 0.007$ &  $N(0.291,0.006^2)$ \\
				Quadratic LD $u_{1, {\rm Johnson-}B}$   & $0.580 \pm 0.027$ &  $N(0.543, 0.026^2)$ \\
				Quadratic LD $u_{2, {\rm Johnson-}B}$   & $0.022 \pm 0.020$ &  $N(0.209, 0.019^2)$ \\
				Quadratic LD $u_{1, \rm Blue-Blocking}$ & $0.275 \pm 0.014$ &  $N(0.276, 0.015^2)$    \\
				Quadratic LD $u_{2, \rm Blue-Blocking}$ & $0.298 \pm 0.006$ &  $N(0.298, 0.064^2)$     \\
				Quadratic LD $u_{1, \rm TESS}$          & $0.254 \pm 0.016$ &  $N(0.252, 0.015^2)$ \\
				Quadratic LD $u_{2, \rm TESS}$          & $0.296 \pm 0.008$ &  $N(0.295, 0.005^2)$ \\
				\hline
		\end{tabular}}%}
		\caption*{\textbf{Supplementary Table 5.} Priors used for WASP-193 for final global join analysis.}
		\label{tableLD}
	\end{center}
\end{table*}

\clearpage

%%%% Figures %%%%%

\begin{figure*}[!ht]
	\centering
	\includegraphics[scale=0.5]{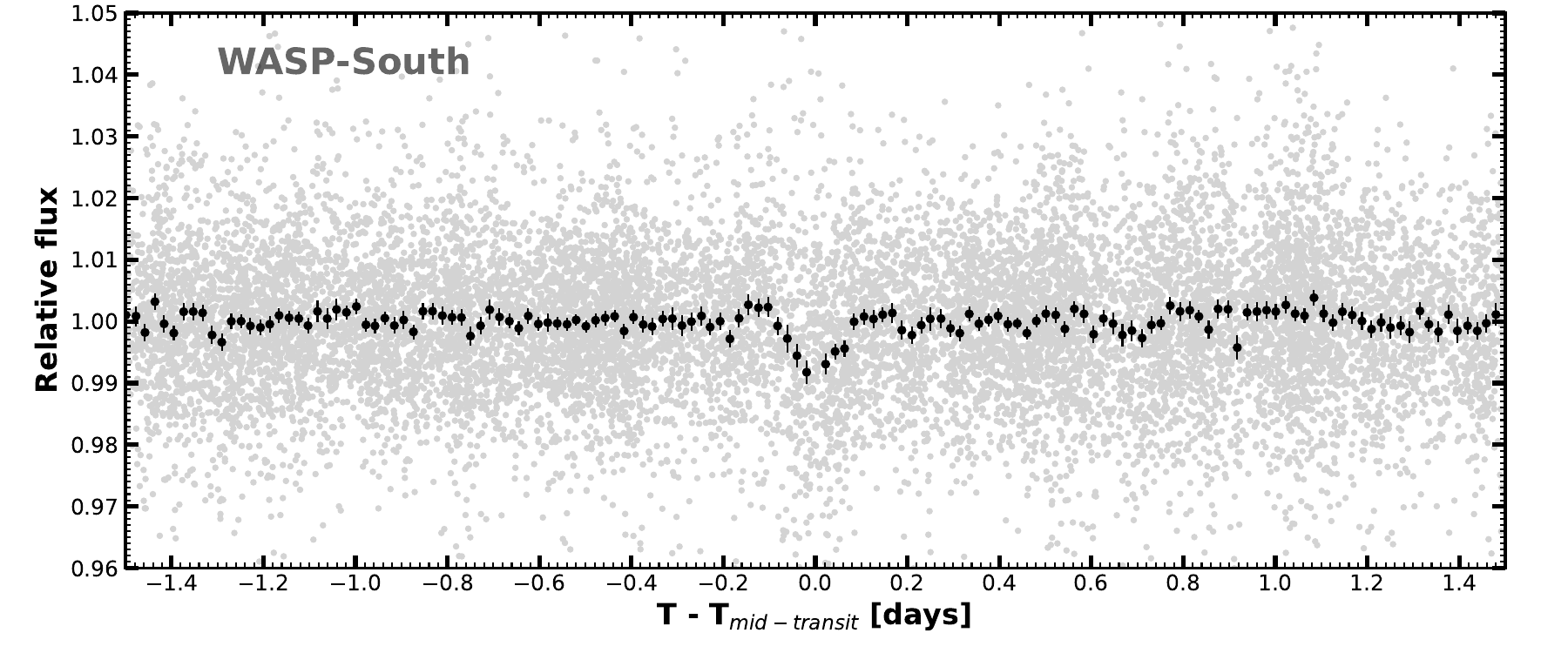}
	\caption*{\textbf{Supplementary Figure 1.} Detrended discovery light-curve of WASP-193 obtained with WASP-South (gray points: unbinned and black points: bin width = 30~min), period-folded using the orbital period deduced from our data analysis.}
	\label{wasp_lc}
\end{figure*}

\begin{figure*}[!ht]
	\centering
	\includegraphics[scale=1.2]{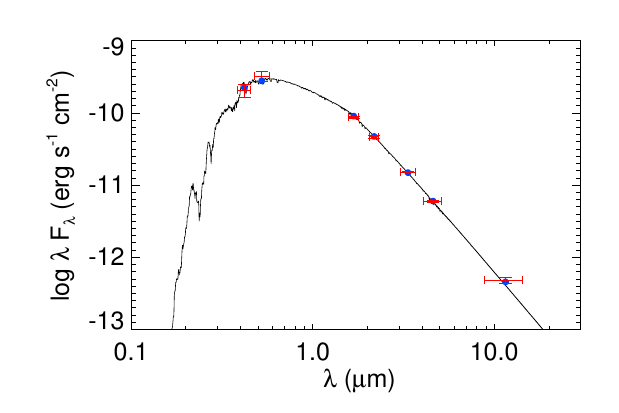}
	\caption*{\textbf{Supplementary Figure 2.} SED model (black line) of WASP-193 and the {\it EXOFASTv2} fit with broads and averages (blue circles) and broadband measurements (red). The error bars in wavelength denote the bandwidth of the corresponding filter and the error bars in flux denote the measurement uncertainties. }
	\label{SEDplot}
\end{figure*}

\begin{figure*}[!ht]
	\centering
	\includegraphics[scale=0.5]{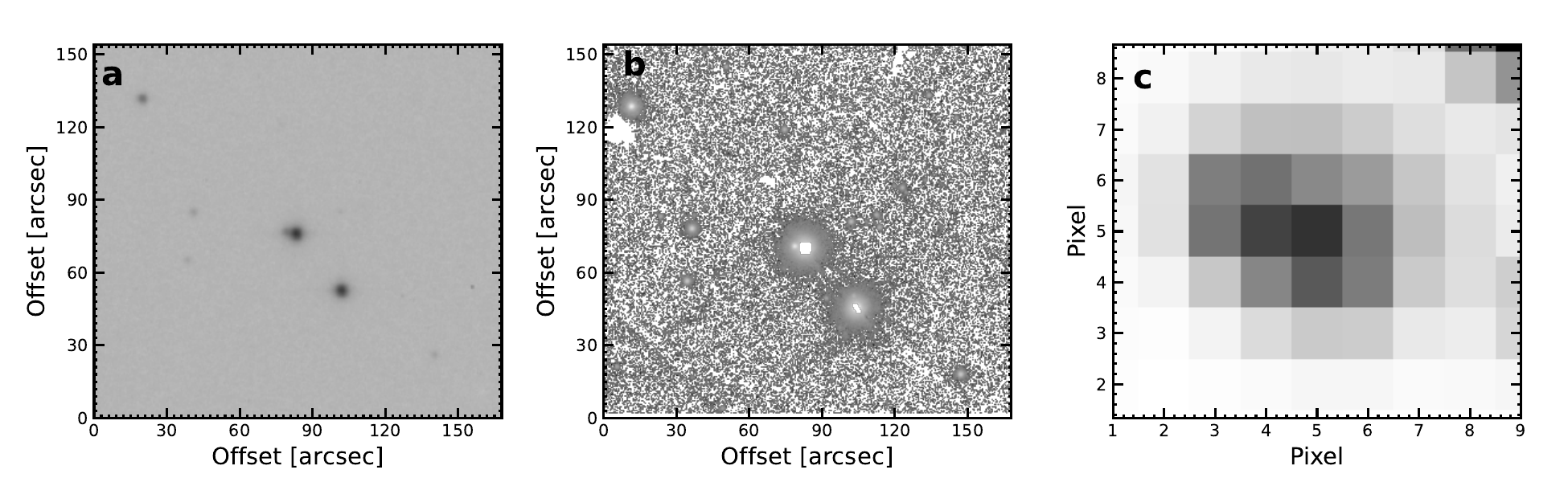}
	\caption*{\textbf{Supplementary Figure 3.} {\bf a}, TRAPPIST-South $z'$ image cropped on a  150$\times$150'' region around WASP-193. {\bf b}, Pan-STARRS\cite{Chambers_2016_Pan-STARRS} $z'$ image of the same 150$\times$150'' region around  WASP-193 with an image-scale of 0.25''~pixel$^{-1}$. {\bf c}, \emph{TESS} 150$\times$150'' image with the same FOV as the panels $b$ and $c$.}
	\label{wasp193_fov}
\end{figure*}

\begin{figure*}[!ht]
	\centering
	\includegraphics[scale=0.4]{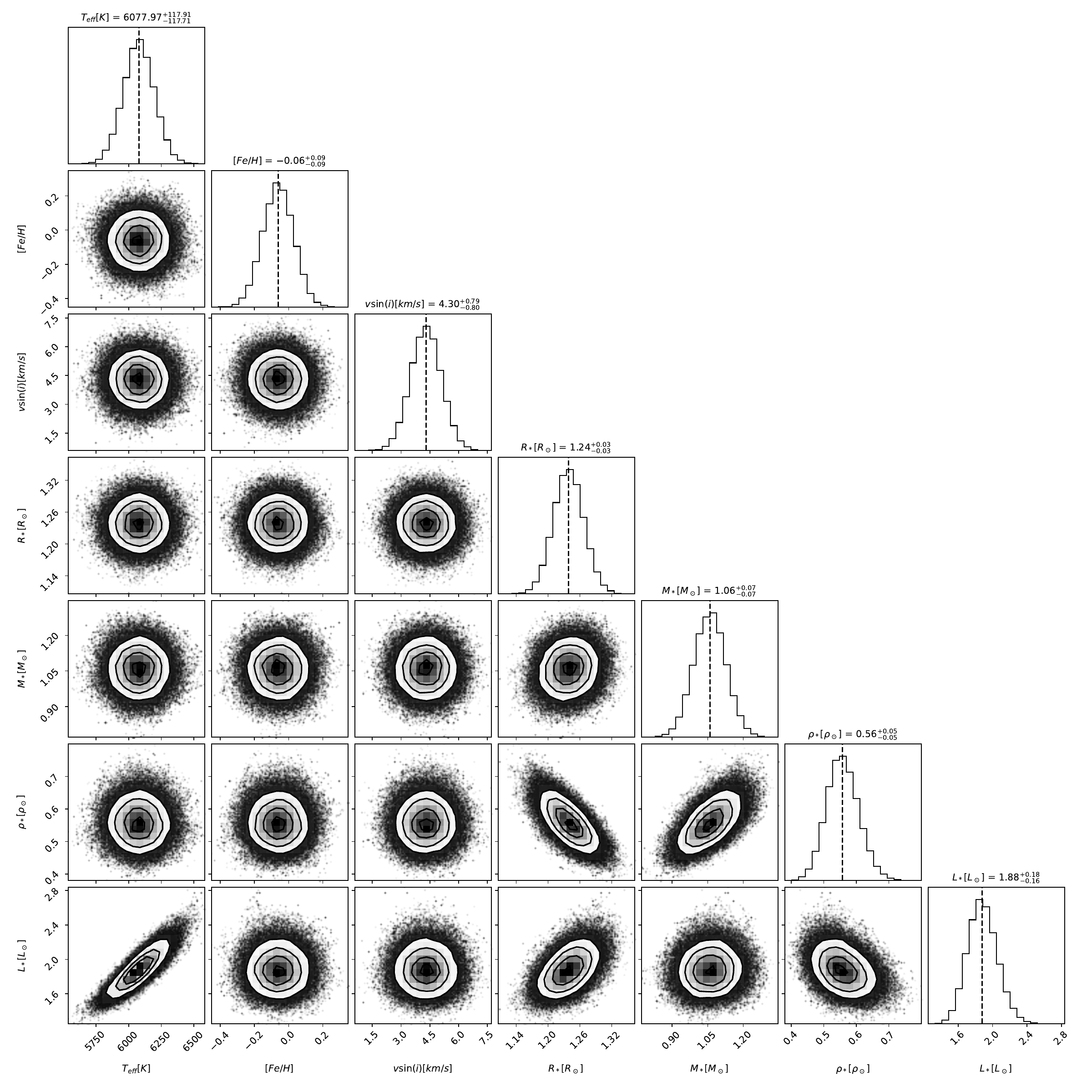}
	\caption*{\textbf{Supplementary Figure 4.} Posterior probability distribution for the stellar physical parameters. These parameters were fitted using our MCMC code as described in Methods. The vertical lines present the median value.}
	\label{corner_star}
\end{figure*}

\begin{figure*}[!ht]
	\centering
	\includegraphics[scale=0.22]{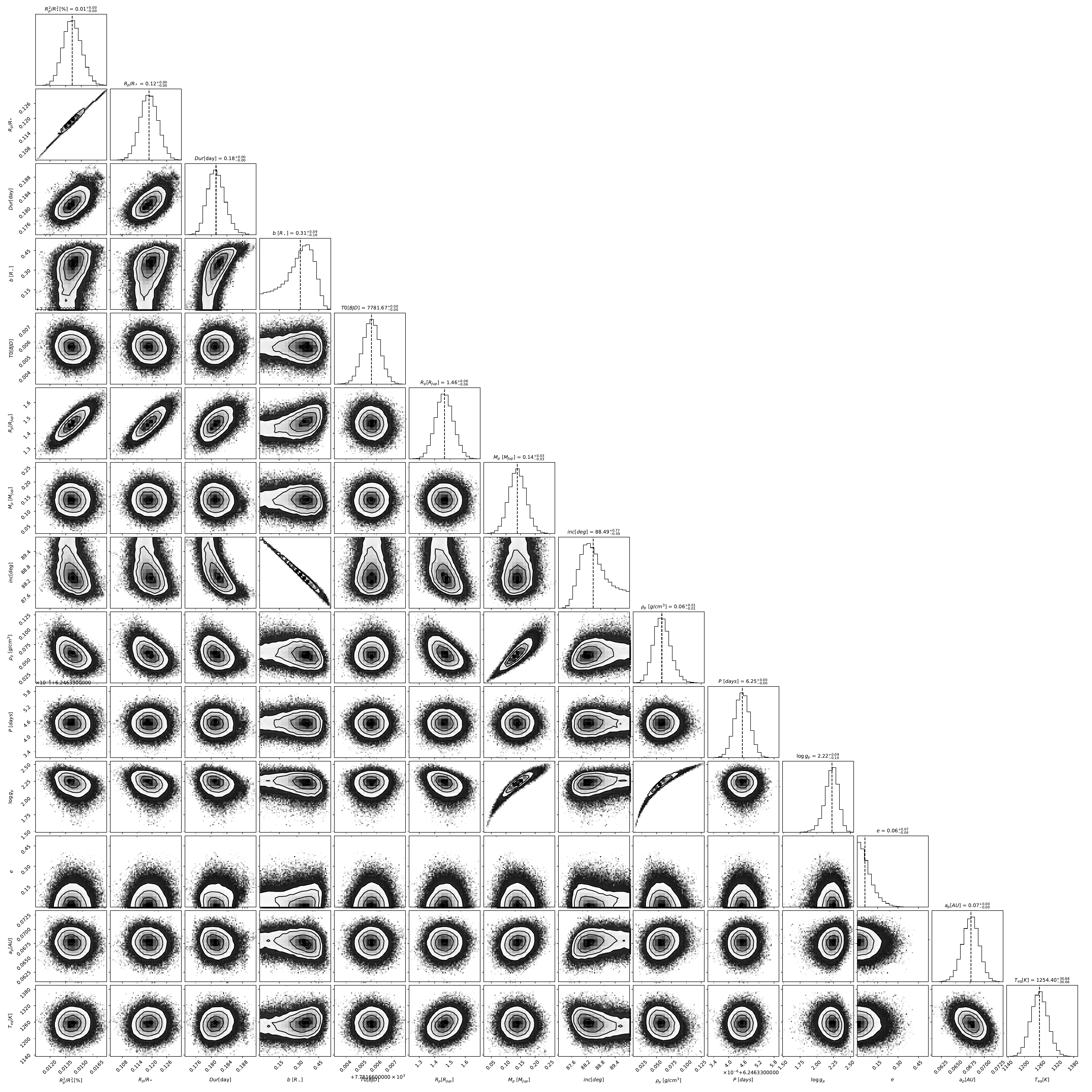}
	\caption*{\textbf{Supplementary Figure 5.} Posterior probability distribution for the planetary physical parameters. These parameters were fitted using our MCMC code as described in Methods. The vertical lines present the median value.}
	\label{corner_planet}
\end{figure*}

\pagebreak
%%% Biblio
\section*{References}
% \bibliographystyle{naturemag}
% \bibliography{Wasp-193_bib}

\end{document}